\documentclass[11pt,preprint]{aastex}

\newcommand{\Msun}{\ensuremath{{\rm M}_{\sun}}}

\newcommand{\iso}[2]{\hbox{${}^{#1}{\rm #2}$}}

\newcommand{\myemail}{akarakas@ap.smu.ca}

\shorttitle{The Production of Mg in AGB Stars}

\shortauthors{Karakas et al.}

\received{2005 July 7}
\begin{document}


\title{The Uncertainties in the \iso{22}Ne $+ \, \alpha$-capture Reaction Rates
and the Production of the Heavy Magnesium Isotopes in Asymptotic Giant Branch Stars
of Intermediate Mass}


\author{A. I. Karakas\altaffilmark{1,2,3}}
\affil{Institute for Computational Astrophysics, Department of Astronomy \& Physics,
Saint Mary's University, Halifax, NS, B3H 3C3, Canada}
\email{\myemail}

\author{M. A. Lugaro\altaffilmark{4}}
\affil{Institute of Astronomy, University of Cambridge,
Madingley Road, Cambridge CB3 0HA, UK}
\email{mal@ast.cam.ac.uk}

\author{M. Wiescher, J. G\"orres and C. Ugalde\altaffilmark{5}} 
\affil{Joint Institute for Nuclear Astrophysics, 
Department of Physics, University of Notre Dame, Notre Dame, IN 46556}
\email{mwiesche@nd.edu, goerres.1@nd.edu, ugalde.1@nd.edu}


\altaffiltext{1}{Swinburne Centre for Astrophysics 
and Supercomputing, Swinburne University, Mail \#31, PO Box 218, Hawthorn, 
Victoria, 3122, Australia}
\altaffiltext{2}{Centre for Stellar \& Planetary Astrophysics, Monash University,
Clayton VIC 3800, Australia}
\altaffiltext{3}{present address: Origin's Institute, Department of 
Physics \& Astronomy, McMaster University, Hamilton ON, Canada}
\altaffiltext{4}{present address: Astronomical Institute, Princetonplein 5,
3584 CC Utrecht, The Netherlands}
\altaffiltext{5}{present address: Department of Physics and Astronomy, 
University of North Carolina, CB 3255, Chapel Hill, NC 27599 and Triangle 
Universities Nuclear Laboratory, Duke University, CB 90308, Durham, NC 27708}

\begin{abstract}

\par
We present new rates for the \iso{22}Ne$(\alpha, n$)\iso{25}Mg
and \iso{22}Ne$(\alpha,\gamma)$\iso{26}Mg reactions, with uncertainties
that have been considerably reduced compared to previous estimates, and
we study how these new rates affect the production of the heavy
magnesium isotopes in models of intermediate mass Asymptotic Giant Branch
(AGB) stars of different initial compositions. 
All the models have deep third dredge-up, hot bottom burning and mass loss. 
Calculations have been performed using the two most commonly used estimates 
of the \iso{22}Ne $+ \, \alpha$ rates as well as the new recommended rates, 
and with combinations of their upper and lower limits. 
The main result of the present study is that with the new rates, uncertainties 
on the production of isotopes from Mg to P coming from the 
\iso{22}Ne $+ \alpha$-capture rates have been considerably reduced. 
We have therefore removed one of the important sources of uncertainty to 
effect models of AGB stars.  We have studied the effects of varying
the mass-loss rate on nucleosynthesis and discuss other 
uncertainties related to the physics 
employed in the computation of stellar structure, such as the modeling 
of convection, the inclusion of a partial mixing zone and the definition 
of convective borders.
These uncertainties are found to be much larger than those coming from
\iso{22}Ne $+ \, \alpha$-capture rates, when using our new estimates.
Much effort is needed to improve the situation for AGB models.

\end{abstract}




\section{Introduction}

\par
The origin of the stable magnesium isotopes, \iso{24}Mg, \iso{25}Mg 
and \iso{26}Mg, is of particular interest to astrophysics because 
Mg is one of the few elements for which we can obtain isotopic information
from stellar spectroscopy.  The ratio \iso{24}Mg:\iso{25}Mg:\iso{26}Mg has 
been derived from high-resolution spectra of cool dwarfs and giants in
the thin and thick disk of the Galaxy \citep*{gl00,yong03b}, and for
giants stars in the globular cluster (GC) NGC 6752 \citep{yong03a}.  
These observations show that many
of the stars, including relatively metal-poor stars ([Fe/H] $\lesssim -1.0$),
have non-solar Mg isotopic ratios\footnote{The solar Mg isotopic ratios are
\iso{24}Mg:\iso{25}Mg:\iso{26}Mg = 79:10:11 \citep{lodder03}.} 
with enhancements in the neutron-rich isotopes, \iso{25}Mg and \iso{26}Mg, 
compared to what is expected from galactic chemical evolution (GCE) models.  
The main stellar nucleosynthesis site for all three stable 
isotopes is hydrostatic burning in the carbon and neon shells of 
massive stars that explode as Type II supernovae \citep{ww95,cl04}.
The abundances of the neutron-rich Mg isotopes are further  
enhanced by secondary $\alpha$-capture processes 
operating in the helium shell.
The amount of \iso{24}Mg produced does not strongly depend on the 
initial metallicity of the model and is an example of primary
nucleosynthesis\footnote{Primary production means 
that the species is produced from the hydrogen and helium initially 
present in the star and the amount produced is relatively independent 
of the metallicity, $Z$, whereas secondary production requires 
some heavier seed nuclei to be present, and the amount produced 
scales with $Z$.}, whereas the amounts of \iso{25}Mg and \iso{26}Mg
produced scale with the initial metallicity of the star.
This means that the Mg content of the  ejecta of low metallicity 
supernovae will be mostly \iso{24}Mg with very little 
\iso{25}Mg and \iso{26}Mg produced, with typical ratios 
\iso{24}Mg:\iso{25}Mg:\iso{26}Mg $\approx$ 99.0:0.50:0.50 from a 25$\Msun$
supernova model with metallicity $Z=0.01 Z_{\odot}$  \citep{ww95}.
\par
Previous studies have shown that GCE models
using ejecta from massive stars match the observational data well for 
[Fe/H] $> -1.0$ but severely underestimate \iso{25,26}Mg/\iso{24}Mg 
at lower metallicities \citep*{timmes95}, indicating that another
production site for the neutron-rich Mg isotopes at low metallicities 
is required to account for the observations.  Recently, \citet{fenner03} 
included the predicted stellar yields of \iso{25}Mg and \iso{26}Mg from 
Asymptotic Giant Branch (AGB) stars \citep{kara03} along with 
yields from Type II supernovae \citep{ww95,lc02} into a GCE model 
of the solar neighborhood.  The GCE model with the AGB contribution 
could successfully match the Mg isotopic ratios of the metal-poor
Galactic disk stars while the model without an AGB contribution
could not. This result indicates that low-metallicity intermediate
mass AGB stars may play an important role in the production
of these species in galaxies and stellar systems.  
The production of the neutron-rich isotopes in AGB stars is
also of interest in relation to the non-solar Mg isotopic ratios 
observed in giant stars in globular clusters \citep{yong03a,yong05}.
The non-solar Mg isotopic ratios observed in NGC 6752 have
been attributed to AGB stars, but \citet{fenner04} used a 
GCE model with tailor made AGB yields from \citet{campbell05} 
and failed to match the abundance patterns observed in stars in 
that cluster. As \citet{ventura05a,ventura05b}
have pointed out, there are still many uncertainties that effect
the stellar yields, so an AGB solution to the
globular cluster anomalies cannot be ruled out at present. 
\par
Further motivation for the study of the production of the 
Mg isotopes in AGB stars is given by their relevance in the important
current debate on the apparent variation of 
the fine-structure constant \citep*{murphy01,ashen04,fenner05},
and the origin of pre-solar spinel grains, 
some of which show enhancements in both \iso{25}Mg and 
\iso{26}Mg compared to solar \citep{zinner05}.
\par
Briefly, intermediate-mass stars (initial mass $\sim$4 to 8$\Msun$) will 
enter the thermally-pulsing phase with a hydrogen (H)-exhausted core 
mass (hereafter core mass) $M_{\rm c} \gtrsim$ 0.8$\Msun$, after 
experiencing the second dredge-up (SDU) \citep{latt96,bgw99,herwig05}.
The SDU brings the products of H-burning to 
the stellar surface (mostly \iso{4}He and \iso{14}N), and will 
slightly alter the composition of the Mg isotopic ratios, with 
an enrichment in \iso{26}Mg at the expense of \iso{25}Mg.
For the massive, $Z = 0.0001$ models, the SDU is the first time the 
surface abundances are altered, because there is essentially 
no first giant branch phase \citep{herwig04a}.
Following the SDU, the He-burning shell becomes thermally unstable 
and flashes every few thousand years or so. 
The energy from the thermal pulse (TP) drives a convective 
pocket in the He-rich intershell, which thoroughly mixes the
products of He-nucleosynthesis within this region. 
Following the TP, the convective envelope moves inward in mass and 
may reach the region previously mixed by the flash-driven convective 
pocket. This mixing event is known as the third dredge-up (TDU) and 
if it occurs, is responsible for enriching the envelope in material
from the H-exhausted core.  Following the TDU  the star contracts 
and the H-shell is re-ignited and provides nearly all of the surface 
luminosity for the next interpulse period.  The thermal pulse -- 
TDU -- interpulse cycle may occur many times during the TP-AGB phase; 
how many times depends on a number of factors including the 
convective model which determines the surface luminosity and mass-loss
rate and hence the total AGB lifetime \citep{ventura05a}.
\par
Hot bottom burning can also occur when the base of the convective 
envelope becomes hot enough to sustain proton-capture nucleosynthesis 
\citep{latt96}. If temperature at the base of the envelope is 
sufficiently hot (over $\sim 60 \times 10^{6}$\,K),
the NeNa and MgAl chains may operate alongside the CNO cycle; \iso{7}Li 
production is also possible via the Cameron-Fowler mechanism 
\citep{sb92,latt96}, which operates at lower temperatures (typically
$\sim 30-40 \times 10^{6}\,$K).  
\citet{frost98} noted that intermediate-mass 
AGB stars may become luminous, optically obscured carbon stars near the 
end of the TP-AGB, when mass loss has removed much of the envelope, 
extinguishing HBB but allowing dredge-up to continue.
\par
\citet{kara03} described in detail the various nucleosynthesis processes that
alter the Mg isotopic ratios in AGB stars. To summarize, \iso{25}Mg and 
\iso{26}Mg are synthesized in the He-shell during thermal pulses by the reactions
\iso{22}Ne($\alpha,n$)\iso{25}Mg and \iso{22}Ne($\alpha,\gamma$)\iso{26}Mg,
when the temperature exceeds about 300$\times 10^{6}\,$K. 
The amount of Mg produced depends on the thermodynamic conditions inside 
the pulse as well as on the composition of the intershell, which will have 
been altered by previous H and He-burning. Neutron captures, 
in particular the  \iso{25}Mg($n, \gamma$)\iso{26}Mg reaction, can also alter 
the Mg isotopic ratio in the intershell, where the neutrons 
come from the \iso{22}Ne($\alpha, n$)\iso{25}Mg reaction \citep{herwig04b}. 
HBB can also significantly alter the surface Mg isotopic ratio 
via the activation of the MgAl chain, which can result in the destruction
of \iso{24}Mg if the temperature exceeds $\sim 90 \times 10^{6}\,$K.
\par
The stellar yields of \iso{25,26}Mg presented in \citet{kara03} and shown in 
figure~\ref{fig:mg-previous} were calculated
from models covering a range in mass (1$\Msun$ to 6$\Msun$) and
metallicity ([Fe/H] $= 0, -0.3, -0.7$). From this figure we see that
the most massive AGB models produce the most \iso{25,26}Mg, as a consequence of 
higher temperatures in the He-shells compared to lower mass stars; we also
notice an increase in production at a given mass with a 
decrease in metallicity.  The computationally
demanding nature of AGB models precluded a detailed study in that paper 
of the effect of the major uncertainties (mass loss, nuclear
reaction rates, convection).
A recent comprehensive study by \citet{ventura05a,ventura05b} demonstrated
that the predictive power of AGB models is still seriously undermined by
these uncertainties.
The theory of convection has a significant effect
on the structure and nucleosynthesis \citep{ventura05a}, 
whilst varying the mass-loss rate results in larger changes to the stellar
yields than varying the nuclear reaction rates \citep{ventura05b}.
However, the magnitude of the errors associated with the relevant
nuclear reaction rates is still one of the key questions 
concerning predictions of magnesium production in intermediate-mass AGB
models.  Whilst the rates of the 
\iso{14}N($\alpha,\gamma$)\iso{18}F and 
\iso{18}O($\alpha,\gamma$)\iso{22}Ne reactions are well determined
\citep{gorres00,dab03}, the two key $\alpha$-capture reaction rates:
\iso{22}Ne($\alpha,n$)\iso{25}Mg and \iso{22}Ne($\alpha,\gamma$)\iso{26}Mg,
suffer from large uncertainties at  the stellar energies appropriate for 
AGB stars \citep{koehler}.  For example, at typical He-shell burning 
temperatures, $T \approx 300 \times 10^{6}\,$K, the NACRE compilation 
\citep{nacre} give an upper limit  to the \iso{22}Ne($\alpha,n$)\iso{25}Mg 
reaction rate that is about 47 times larger than 
the recommended rate. At lower temperatures, the uncertainties are 
even larger. 
\par  
The aims of this paper are two fold. First, we present new reaction rates for
the two key $\alpha$-capture reactions, with considerably reduced
uncertainties compared to those given by the NACRE compilation. Second,
we use these new rates within models of different metallicities 
([Fe/H] $\approx 0, -0.3, -0.7, -2.3$), for a typical mass (5$\Msun$) 
that produces the Mg isotopes during the thermally-pulsing AGB (TP-AGB) phase.
For each model we calculate the stellar yields and compare to previous
nucleosynthesis calculations using older estimates of these 
reaction rates, including those by \citet{herwig04b} and 
\citet{ventura05a,ventura05b}. We also examine the effect of 
other model uncertainties, in particular the inclusion of a partial
mixing zone and mass loss, on the stellar yields. 
\par
The paper is organized as follows. In \S\ref{section:method} we 
discuss the numerical method used for the stellar model calculations,
including a discussion of the input physics used. In \S\ref{section:rates} 
we present new rates for the \iso{22}Ne $+ \alpha$  capture reactions.  
The results of the calculations are presented in \S\ref{section:results} 
and discussed in \S\ref{section:discussion}. 

\section{The numerical method} \label{section:method}

\par
We calculate the structure first and perform detailed nucleosynthesis calculations
afterward, using a post-processing algorithm. The stellar structure models were 
calculated with the Monash version of the Mount Stromlo Stellar Structure Program; 
see \citet{frost96} and references therein for details.  Mass loss on the 
first giant branch is included using the Reimer's mass-loss prescription
\citep{r75} with the parameter $\eta = 0.4$;  on the AGB we use the formulation
given by \citet{vw93} in all models unless indicated otherwise. 
We calculate two models (5$\Msun$, $Z=0.02$ and $Z=0.0001$) using Reimer's 
mass loss on the AGB with the parameter $\eta = 3.5$. Here we assume that 
the \citet{vw93} and \citet{r75} mass-loss prescriptions, derived for
solar-like metallicities or those appropriate for the Magellanic Clouds,
can be applied to $Z=0.0001$ intermediate-mass AGB models without modification.
The use of either mass-loss law results in a large number of thermal pulses
(up to $\sim 140$).
\par
All models were calculated from the zero-age main sequence to near the end of 
the TP-AGB phase.
The occurrence of the third dredge up and hot bottom burning (HBB) depend critically
upon the convection model used \citep{frost96,mowlavi99,ventura05a} and the method for
determining convective borders.  Briefly, we use the standard 
mixing-length theory (MLT) for convective regions, with a mixing-length 
parameter $\alpha = l / H_P = 1.75$, and determine the border by applying the 
Schwarzschild criterion. Hence we do not include convective overshoot, 
in the usual sense.  We do however, search for a neutral border to the 
convective zone, in the manner described in detail by \citet{latt86,frost96} and 
\citet{kara02}. We note that this method has been shown to increase 
the efficiency of the third dredge-up compared to models that strictly use the 
Schwarzschild criterion.  
\par
We performed detailed nucleosynthesis calculations using a post-processing 
code which includes 74 species and time-dependent diffusive mixing in all 
convective zones \citep{cannon93}. The details of the 
nucleosynthesis network are outlined in \citet{lugaro04}, but we remind the
reader that we include 59 light nuclei and 14 iron-group species. 
We also add the fictional particle $g$, to count the number of neutron
captures occurring beyond \iso{62}Ni  \citep{latt96,lugaro04}.
Initial abundances are taken from \citet{ag89} for the $Z=0.02$ models and we
assume scaled-solar for the $Z=0.0001$ models. We assume an initially 
$\alpha$-enhanced mixture typical of thin-disk stars \citep{reddy03} for the
5$\Msun$, $Z=0.008, 0.004$ models.  The initial Mg isotopic ratios
(e.g. \iso{25}Mg/\iso{24}Mg) are 21.9\% and 35.2\% less than solar in the
$Z=0.008$ and $Z=0.004$ models, respectively.
\par
The bulk of the 506 reaction rates are from the REACLIB data tables \citep{reaclib},
based on the 1991 updated version.  Some of the proton, $\alpha$ and neutron capture
reaction rates have been updated according to the latest experimental results, see 
\citet{lugaro04} for details, and we calculate one set of models using these rates.
We define this set to be our {\em standard} set and the results
from these computations are shown in figure~\ref{fig:mg-previous}.
In all the other sets of computed models we have updated the proton capture rates
for the NeNa and MgAl chains to those recommended by NACRE (see table~\ref{tab:pcaptures}
in \S\ref{app:rates} for details of which proton capture rates that have been updated). 
With this new set of updated rates, we study the effect of changing
the \iso{22}Ne $+ \alpha$ rates: we run models with our standard choice for these
rates \citep[][without the inclusion of the 633 KeV resonance, indicated as K94
in table~\ref{tab:mgyields}]{kaep}, with
the NACRE recommended rates (indicated as NACRE in table~\ref{tab:mgyields}),
and with our new recommended rates, as well as using combinations of the new
upper and lower limits.


\section{Nuclear structure of \iso{26}Mg and the uncertainties in
the \iso{22}Ne + $\alpha$ reaction rates.} \label{section:rates}

The reaction rates for the \iso{22}Ne+$\alpha$ fusion processes are
determined by the level structure of the compound nucleus
$^{26}$Mg above the $\alpha$ threshold at $T_{\alpha}$=10.615 MeV
and near the neutron threshold $T_n$=11.093 MeV. There is only
very limited experimental and theoretical information available
about possible natural parity resonances in that energy range and
both the $^{22}$Ne($\alpha,\gamma$)$^{26}$Mg as well as the
$^{22}$Ne($\alpha,n$)$^{25}$Mg reaction rates have substantial
uncertainties in the temperature range of stellar helium burning.
These uncertainties have been a matter of debate for quite some time.

The uncertainty for the $^{22}$Ne($\alpha,n$)$^{25}$Mg reaction is
mainly determined by the possible contribution of a low energy
1$^-$ resonance at 0.538 MeV which has been observed both in
photon induced neutron emission \citep{berman} as well as in
neutron capture measurements \citep{weigmann, koehler}. Yet it is
not clear how strongly this level is populated in the $\alpha$
channel. Direct $^{22}$Ne($\alpha,n$)$^{25}$Mg measurements
\citep{harms,drotleff,jaeger} were performed over the entire energy
range down to the neutron threshold, but these measurements were
handicapped by cosmic ray induced background in the neutron
detectors, therefore only an upper limit for the resonance
strength was obtained. Also in a $^{22}$Ne($^6$Li,d)$^{26}$Mg
$\alpha$ transfer measurement \citep{giesen} only an upper limit
for the $\alpha$ spectroscopic factor $S_{\alpha}\le$0.2 was
determined. The NACRE compilation \citep{nacre} has based the upper
limit for the reaction rate on the direct $^{22}$Ne($\alpha,n$)
measurement by \citet{drotleff}. For the lower
limit any contribution of this resonance was neglected. While more
recent $^{22}$Ne($\alpha,n$) studies have reduced the upper limit
substantially \citep{jaeger} the proposed value is still
substantially higher than suggested by the $\alpha$ transfer data.
A recent systematic re-analysis by \citet{koehler} of the
$^{25}$Mg($n,\gamma$) measurement by 
\citet{weigmann} pointed out on the basis of a careful R-matrix
analysis of $^{25}$Mg($n, \gamma$)$^{26}$Mg data that at least
three or more additional natural parity states could be expected
above the neutron threshold. It was shown by \citet{koehler} that
in particular the possible contribution of the 2$^+$level at
11.112 MeV to the reaction rate does raise its uncertainty by more
than an order of magnitude. In view of this debate and in absence
of new data it seems worthwhile to review the present status and
suggest more reliable limits for the rate taking into account all
of the presently available experimental information.

\subsection{The reaction rate of $^{22}$Ne($\alpha,n$)$^{25}$Mg}
                                                                                            
As it has been outlined by \citet{koehler}, several natural
parity resonances near the neutron threshold can contribute
significantly to the $^{22}$Ne($\alpha,n$)$^{25}$Mg reaction rate
increasing the neutron production at low energies. The upper
limits given by \citet{koehler} were scaled to the
experimental upper limits of the resonance state at
$E_{\alpha}^{cm}$=0.538 MeV ($E_x$=11.153 MeV, $J^{\pi}$=1$^-$) by
\citet{jaeger}. As pointed out above $\alpha$ transfer
measurements suggest a substantially lower $\alpha$ strength for
the $E_x$=11.153 MeV level. While the experimental lower limit
indicates a resonance strength of $\omega\gamma\le$60 neV the
transfer data reduce the upper limit considerably to
$\omega\gamma\le$27 neV \citep*[see also][]{kaep}
after renormalization of the reference resonance.
                                                                                                        
Based on the data by \citet{giesen} and \citet{koehler}
we have reanalyzed the strength of the low energy
resonances and derived the resonance strength from the $\alpha$
spectroscopic factors normalized to the newly determined strength
of the well known resonance at $E_{\alpha}^{cm}$=0.713 MeV
($E_x$=11.328 MeV, $J^{\pi}$=1$^-$) \citep{jaeger}. The procedure
has been outlined in \citet{giesen}. Table~\ref{(a,n)} 
summarizes the level parameters adopted for
determining the strengths as well as the upper and lower limits
for the neutron unbound states between 11.1 and 11.4 MeV in
$^{26}$Mg. While neutron capture data \citep{weigmann,koehler}
indicate a large number of natural parity states between 11.16 and
11.3 MeV excitation energy, no appreciable $\alpha$
strength has been observed for these levels in the
$^{22}$Ne($^6$Li,d)$^{26}$Mg transfer reaction.  For these states
levels we adopt an average spectroscopic strength of
C$^2$S$_{\alpha}$=0.005. For the two levels at 11.113 and 11.153
MeV at the neutron threshold the experimental data indicate some
$\alpha$ strength, which however could not be reliably determined
due to the lack of experimental resolution in the deuteron
spectrum. We extracted S$_{\alpha}$ spectroscopic factors of 0.005
and 0.01, respectively with an upper limit of 0.01 and 0.02. For
the neutron and $\gamma$-partial widths we adopted the values
given by \citet{koehler} for calculating the resonance
strengths. Using these resonance strengths listed in table
\ref{ogres1} we have calculated the reaction rate contributions of
the resonances for $^{22}$Ne($\alpha,n$)$^{25}$Mg.
\\
The reaction rate is shown as a function of temperature in figure
\ref{22ne+a}. In the temperature range between 0.3 and 0.5~GK %
the reaction rate is dominated by the contribution of the lowest
measured resonance at E$_{\alpha}^{cm}$=0.713 MeV \citep{jaeger}. At
higher temperatures the rate is determined by higher energy
resonance contributions resonances \citep{harms,drotleff}. The
threshold resonance states determine the rate at lower
temperatures $\le$ 0.3 GK. The largest uncertainty is obviously
associated with this low temperature range. Based on the presently
available data the reaction rate is uncertain by roughly one order
of magnitude. This uncertainty depends mainly on the uncertainty
in the spectroscopic $\alpha$ strengths of the threshold
resonances. More detailed experimental work is necessary to reduce
the present uncertainties.

\subsection{The reaction rate of \iso{22}Ne($\alpha,\gamma$)\iso{26}Mg}

It has been argued before \citep{kaep} that the
\iso{22}Ne($\alpha,\gamma$)\iso{26}Mg
reaction may compete strongly
at low temperature with the \iso{22}Ne($\alpha,n$)\iso{25}Mg
capture process. This possibility depends sensitively on the
contribution of resonances at energies below the neutron
threshold. Koehler has argued that the strong
resonance at $E_{\alpha}^{cm}$=0.711 MeV ($E_x$=11.326 MeV,
$J^{\pi}$=1$^-$), which was observed by \citet{wolke}
does not correspond to the resonance level observed in the
competing $^{22}$Ne($\alpha,n$) reaction channel
\citep{harms,drotleff,giesen,jaeger}. Only resonance levels with
small neutron widths $\Gamma_n\le\Gamma_{\gamma}$ can have an
appreciable resonance strength in the \iso{22}Ne($\alpha,\gamma$)
channel.

We have recalculated the resonance strengths $\omega\gamma$ of all
possible resonances in the $^{22}$Ne($\alpha,\gamma$) channel. The
resonance strengths scales directly with the spectroscopic factors
S$_{\alpha}$ \citep{giesen}, which have been re-normalized to the
experimentally determined strength of the state at 11.328 MeV
\citep{jaeger}. The spin parity assignment for the
low energy sub-neutron threshold resonances is based on the DWBA
analysis of the $^{22}$Ne($^6$Li,d)$^{26}$Mg data. The most
important resonance is the one at $E_{\alpha}^{cm}$=0.33 MeV
($E_x$=10.945 MeV, $J^{\pi}$=2$^+$,3$^-$). The data do not allow a
unique spin assignment for this level; this uncertainty represents
the main uncertainty in the spectroscopic factor and resonance
strength. These parameters were used to estimate the resonance
strengths for the not yet observed low energy resonances; the
results are listed in Table~\ref{ogres2}.

The reaction rate for $^{22}$Ne($\alpha,\gamma$)$^{26}$Mg is shown
in figure~\ref{22ne+a} as a function of temperature. For
temperatures below 0.3 GK the resonance at $E_{\alpha}^{cm}$=0.33
MeV clearly dominates the reaction rate, possible contributions
can also come from the two 1$^-$ resonances at
$E_{\alpha}^{cm}$=0.538 MeV ($E_x$=11.153 MeV) and
$E_{\alpha}^{cm}$=0.568 MeV ($E_x$=11.183 MeV) at the neutron
threshold. Based on the recent re-analysis of the neutron capture data 
\citep{koehler} these states are expected to have only small neutron partial
widths. The reaction rate in the temperature range T$\ge$0.3 GK
is determined by the contributions of states above 11.3 MeV, in
particular by the strong level at $E_{\alpha}^{cm}$=0.711 MeV
($E_x$=11.326 MeV) which has been measured by \citet{wolke}. 
Also indicated is the uncertainty range for the
$^{22}$Ne($\alpha,\gamma$)$^{26}$Mg rate. It is obvious from this plot that the
experimental uncertainties associated with the low energy
resonances dominate the uncertainty in the low temperature range.

\subsection{The comparison of the reaction channels}
                                                             
The temperature dependence of the two $^{22}$Ne+ $\alpha$ reaction rates
differ considerably at lower temperatures because the neutron
channel opens only at higher energies. Figure~\ref{22ne+a}
demonstrates that at low temperature T$\le$0.2GK the
($\alpha,\gamma$) reaction channel dominates over the competing
($\alpha,n$) reaction process. However, there are still considerable uncertainties
associated with the low temperature range of the reaction rates as demonstrated in 
figure~\ref{ratio}. Shown are the upper and lower limits of the reaction
rate normalized to the recommended value and clearly demonstrates
that below T$\le$0.3 GK the uncertainty range rapidly increases
towards lower temperatures to one order of magnitude for the
$^{22}$Ne($\alpha,n$)$^{25}$Mg rate and to nearly two orders of
magnitude for the $^{22}$Ne($\alpha,\gamma$)$^{26}$Mg rate. This
affects the reliability for nucleosynthesis predictions in rapidly
changing low temperature environments. Figure~\ref{anag} shows the
ratios of $^{22}$Ne($\alpha,n$)$^{25}$Mg rate
and $^{22}$Ne($\alpha,\gamma$)$^{26}$Mg rate. A comparison between the
upper and lower limits of the respective reaction rates shows the
uncertainty range for this ratio. At higher temperatures the
$^{22}$Ne($\alpha,n$)$^{25}$Mg reaction clearly dominates while towards
lower temperatures the $^{22}$Ne($\alpha,\gamma$)$^{26}$Mg
reaction will be far stronger than the competing $^{22}$Ne($\alpha,n$)$^{25}$Mg
channel. However, the present data are not stringent enough to determine the
exact temperature for this cross over but can be limited to
a temperature range between 0.15 and 0.3 GK. Further measurements
of low temperature resonances in both reaction channels are therefore
necessary to improve on the present uncertainties.

\section{Results} \label{section:results}

We now present the results of the stellar models. Unless otherwise
stated, we discuss the models computed using the \citet{vw93}
mass-loss rates.

\subsection{Structural properties and the CNO nuclei}
\par 
In this section we present some of the structural properties of the
stellar models used for the present study in detail, including a 
brief discussion of the evolution of the CNO nuclei during the TP-AGB phase.
In table~\ref{table:agbmodels} we first include the initial mass and 
metallicity and the mass-loss law adopted for the structure calculation.  
VW93 refers to models calculated with \citet{vw93} mass loss whilst 
R75 refers to models calculated with \citet{r75} with the parameter 
$\eta_{\rm R} = 3.5$.  Furthermore, we present the 
core mass at the beginning of the 
TP-AGB, the mass of the envelope when HBB is shutoff and the mass
at the end of the calculation, the total number of TPs, the number of 
TDU episodes, the total amount of matter dredged into the envelope 
$M_{\rm dred}^{\rm tot}$, the maximum temperature 
at the base of the convective envelope, the maximum temperature 
in the He-shell during a TP and the maximum dredge-up efficiency, 
$\lambda_{\rm max}$. In table~\ref{tab:agb2} we present some surface
abundances results including the final C$+$N$+$O value divided by 
the initial (see table caption for more details).
The dredge-up efficiency is defined by
$\lambda = \Delta M_{\rm dredge}/\Delta M_{\rm h}$, where 
$\Delta M_{\rm dredge}$ is the amount of H-exhausted core matter 
mixed into the envelope and $\Delta M_{\rm h}$ is the amount by 
which the core mass grew during the previous interpulse period. 
For all the models considered here, $\lambda \gtrsim 0.9$,
is reached after a small number ($\sim 6$) of pulses, similar to
behavior observed by \citet{stancliffe04} and \citet{herwig04a}
for the same mass.
\par
The VW93 mass-loss rate depends on the pulsation period until
the period reaches 500\,days, after which the luminosity-driven
superwind phase begins. The pulsation period \citep{vw93} depends in
turn on the radius and total mass and because the lower metallicity models
are more compact they take longer (in time) to reach the start of the
superwind phase and hence experience more TPs than the solar 
metallicity models.
In comparison, the R75 mass-loss rate is proportional to both the radius 
and the luminosity and as we can see from table~\ref{table:agbmodels} 
at the lowest metallicity we consider ($Z = 10^{-4}$) the large luminosity
at this composition results in fewer TPs than using VW93 mass loss, 
the reverse of the behavior observed at $Z=0.02$. 
\par
For all models except the $Z=0.02$ cases, the temperature at the
base of the envelope, $T_{\rm BCE}$, increases quickly at the
beginning of the TP-AGB and exceeds $50 \times 10^{6}\,$K
by about the 8$^{\rm th}$ pulse.   The envelope
mass after which HBB is shutoff decreases with a decrease 
in metallicity, see column~5 in  table~\ref{table:agbmodels}.
For the solar composition VW93 (R75) models, $T_{\rm BCE}$
reaches $50 \times 10^{6}\,$K after about the 9$^{\rm th}$ TP 
(12$^{\rm th}$ TP) and decreases quickly once the envelope mass
drops below $\approx 2.6\Msun$, which is when HBB is shutoff.
After this, dredge-up continues increasing the C/O ratio 
above unity in some cases (see table~\ref{tab:agb2}).
\par 
For the remainder of this section we summarize the nucleosynthesis 
of the CNO nuclei, because these species are not affected by 
varying the \iso{22}Ne $+ \alpha$ reaction rates.  
The contribution of many TPs with efficient dredge-up means that 
there is a considerable increase in the amount of \iso{12}C 
that can be converted to primary \iso{14}N by HBB (note also
the low final \iso{12}C/\iso{13}C ratios shown in 
table~\ref{tab:agb2}). This increase
can be best appreciated by comparing the final envelope C$+$N$+$O 
abundance (divided by the initial) for each model in 
table~\ref{tab:agb2}. In the lowest $Z$ models \iso{14}N
is the most abundant isotope, and an increase in $Z$ results
in decreasing levels of enrichment, where the final 
[\iso{14}N/Fe]\footnote{assuming solar values from
\citet{ag89}, as used in the $Z=0.02$ models. We use the standard
logarithmic notation 
[$X/Y$] $= \log_{10}(X/Y)_{*} - \log_{10}(X/Y)_{\odot}$.}
values are 3.9, 1.9, 1.5 and 0.6 in the $Z=0.0001, 0.004, 0.008$ 
and 0.02 model, respectively. 
[C/N] ratios have been observed in metal-poor stars in the halo, 
with the finding that $-0.5 <$ [C/N] $< 1.5$ at carbon abundances 
around [C/Fe] $\sim 2$ \citep{johnson05}. 
Our models overproduce N compared to C, with [C/N] $< -1$ 
in all metal-poor 5$\Msun$ computations. 
\par
The solar-metallicity VW93 model does not become carbon rich, 
whereas the R75 model becomes a carbon star after the 
28$^{\rm th}$ TP.  From table~\ref{table:agbmodels} we see
that the reason for this is more TDU episodes, resulting in
about a factor of two more matter dredged into the envelope.
The 5$\Msun$, $Z=0.008$ model does not become a carbon star,
whereas the $Z=0.004$ model does. The final [\iso{12}C/Fe] 
values for both cases are 0.16 and 0.70, compared
to 0.09 and 0.16 initially.  The $Z=0.004$ model has a longer 
TP-AGB phase (0.59~Myr compared to 0.50~Myr), leading to more \iso{16}O
destroyed by HBB, even though HBB temperatures are similar in both 
models.  The final [\iso{16}O/Fe] values are 0.02 and 0.04, 
compared to the initial enhanced values of 0.16
and 0.24, respectively.
The 5$\Msun$, $Z=0.0001$ models (with VW93 and R75 mass loss)
become carbon stars quickly, after the second TP. This is a 
consequence of deep third dredge-up and a low initial \iso{16}O 
envelope abundance.  The final [\iso{12}C/Fe] values are 2.4 
and 2.6, in the VW93 and R75 models.  
The envelope \iso{16}O abundance is destroyed during the 
interpulse by HBB at temperatures $\approx 90 \times 10^{6}\,$K 
whereas the many efficient TDUs result in an overall increase,
leading to a final [\iso{16}O/Fe] $\approx 1$ in both cases. 
This behavior was also observed by \citet{herwig04a} for 
the equivalent mass and metallicity.

\subsection{Surface abundance evolution of the Mg isotopes}

The envelope abundances of the Mg isotopes are first
modified by the operation of the SDU, and the magnitude of such
modifications increases with a decrease in the metallicity or an
increase increase in the mass.
For the 5$\Msun$, $Z=0.02$, the envelope \iso{25}Mg abundance decreases 
by 4.2\% while \iso{26}Mg increases by 3.7\% after the SDU, whereas 
in the 5$\Msun$, $Z=0.004$ model we observe a decrease of 9.4\% in the 
\iso{25}Mg abundance and an 8.4\% increase in \iso{26}Mg.
For all cases but the 5$\Msun$, $Z=0.0001$ model, the \iso{24}Mg 
abundance does not change.  For this model, the \iso{24}Mg abundance
increases by 11\%, whereas the \iso{25}Mg and \iso{26}Mg abundances
decrease by 13\% and 6.8\% respectively. 
The results quoted above are for the calculations that use the
NACRE NeNa and MgAl chain reaction rates.
\par
In figure~\ref{m5z02} we show the evolution during the TP-AGB phase 
of the \iso{25}Mg (solid line) and \iso{26}Mg (dashed line) 
abundances at the surface of the 5$\Msun$, $Z=0.02$ model for 
four different choices of the 
\iso{22}Ne $+ \alpha$ reaction rates.  The abundances 
in figure~\ref{m5z02} are the mole fraction, $Y$ (where the mass
fraction $X = YA$ and $A$ is the atomic mass), scaled to the total 
magnesium abundance, Mg $=$ \iso{24}Mg$+$\iso{25}Mg$+$\iso{26}Mg. 
In figure~\ref{mg-surf} we show the surface abundance evolution of
the Mg isotopes for three different initial compositions,
that use the new recommended rates for the \iso{22}Ne $+ \alpha$ reactions.
In table~\ref{tab:agb2} we also present the Mg isotope ratios 
(in terms of \iso{24}Mg:\iso{25}Mg:\iso{26}Mg)
at the beginning and at the end of the TP-AGB phase for each model.
Figure~\ref{mg-surf} and table~\ref{tab:agb2} shows the influence of 
metallicity on the evolution of the Mg isotopes, where the effects of 
HBB become more significant at lower $Z$.
\par
From figure~\ref{m5z02} we see the important result that using the 
new recommended rates
for the \iso{22}Ne $+ \, \alpha$ reactions does not result in large changes
to the production of \iso{25,26}Mg compared to the calculations performed
using older estimates. For the solar metallicity model, the final surface Mg 
isotopic ratios are 70.8:13.5:15.6 when using NACRE compared to 
72.2:12.4:15.4 when using the new recommended rates. The only observable
change is less \iso{25}Mg but about the same amount (or just slightly less)
\iso{26}Mg.  These results  are consistently observed in the metal-poor AGB 
models, although the trends are stronger at lower metallicity.
\par
To explain the above trends, we need to examine the behavior
of the new rates compared to the NACRE and K94 rates.
We produce less \iso{25}Mg because
the older estimates for the \iso{22}Ne($\alpha, n$)\iso{25}Mg 
reaction are up to 40\% faster at typical He-shell burning 
temperatures i.e. $250 \times 10^{6}$K to $400 \times 10^{6}$K. 
Why the production of \iso{26}Mg is reduced or remains about
the same is not immediately obvious given that the new recommended rate
is faster than both the NACRE (by up to 26\%) and K94 (by 70\% to
100\%) rates. The reason is again related to the ($\alpha, n$)
reaction which releases neutrons during a TP, some of which
are captured to produce \iso{26}Mg, a process that is most
efficient in the lowest $Z$ models.
In the solar metallicity models the production of \iso{26}Mg 
is less dependent on the neutron flux and we produce about 
the same when using the new recommended 
rate for the \iso{22}Ne($\alpha,\gamma$)\iso{26}Mg reaction.

\subsection{Stellar yields}
\par
In the first row of table~\ref{tab:mgyields} we present the stellar
yields of \iso{25}Mg and \iso{26}Mg (in $\Msun$) for the 5$\Msun$
models of different $Z$, computed using the new recommended rates.
In other rows we present the percentage difference between yields
calculated with a different set of rates (as noted in the table)
and our reference yields given in the first row\footnote{according 
to [yield$(i) -$ yield(ref)]/yield(ref) multiplied by 100, where 
$i$ is one of the combinations of reaction rates described in 
\S\ref{section:method}}.
In figure~\ref{yield-compare1} we show the percentage
difference between the stellar yields calculated using the new recommended
rates (the reference), the new upper and lower limits and the
NACRE recommended rates, for two different metallicities. 
The $x$-axis is atomic mass, and we present percentage differences for
\iso{22}Ne through to \iso{31}P, with \iso{26}Al included at $26.5$. 
According to the
definition of the percentage difference given above, a positive 
difference means that that model produced (or destroyed) more of 
a given species compared to our reference case. 
To compare to the magnitude of the uncertainties associated the NACRE
rates, in figure~\ref{yield-compare2} we show the percentage 
difference for models that use the NACRE recommended (the reference)
and the NACRE upper and lower limits. We also vary the 
\iso{25}Mg($n, \gamma$)\iso{26}Mg reaction rate by a factor 
of two each way (see figure~\ref{yield-compare2}).
\par
These figures show that using the new rates for the \iso{22}Ne 
$+ \, \alpha$ reactions results in considerably 
smaller uncertainties in the production of all isotopes between \iso{22}Ne to 
\iso{31}P compared to models that use the NACRE rates. 
For the \iso{25,26}Mg isotopes, differences of the order $\sim$400\% in
the case of the 5$\Msun$, $Z=0.02$ model with the NACRE upper limits are
reduced to less than 30\%, when using the new upper limits.  For
models of lower metallicity, the uncertainties are even smaller (at most
$\sim 20$\%).
Varying the \iso{25}Mg($n, \gamma$)\iso{26}Mg reaction
rate given by \citet{weigmann} by a factor of two each way results in 
differences of at most 33\% in the production of \iso{31}P, about 26\% 
in the production of \iso{25}Mg and about 15\% in the production of 
\iso{26}Mg for all models, regardless of metallicity. Note that the
abundance of heavier isotopes are affected by neutron captures hence
the dependence on these reaction rates.
\par 
The percentage differences reported in table~\ref{tab:mgyields}
and shown in figures~\ref{yield-compare1} and~\ref{yield-compare2}
illustrate that the computed errors decrease with a decrease in the
metallicity of the model. The reason could be that the yields
significantly increase with an decrease in $Z$, and hence 
changes to a larger number result in smaller differences than
changes to a smaller number (in the case of the $Z=0.02$
models). Also, the low $Z$ models have more TPs and more efficient 
HBB means that the proton-capture MgAl chain reactions have a 
greater overall effect on the stellar yields. 

\subsection{The effect of partial mixing zones and varying the mass-loss rate}

The inclusion of a partial mixing zone (PMZ) at the deepest penetration 
of the TDU will mix protons from the envelope into the He-intershell, 
producing a \iso{13}C pocket. 
In the PMZ neutrons are liberated during the interpulse period by the
neutron source reaction \iso{13}C($\alpha,n$)\iso{16}O 
and the Mg isotopic abundances can be modified
because of the activation of the chain of neutron captures starting from the
abundant \iso{22}Ne and proceeding through \iso{23}Na to 
\iso{24}Mg, \iso{25}Mg and \iso{26}Mg. 
Given the high value of the neutron capture cross section of \iso{25}Mg,
with respect to those of \iso{24}Mg and \iso{26}Mg \citep[see][]{bao00},
neutron captures in the \iso{13}C pocket produce \iso{24}Mg and 
\iso{26}Mg, while depleting \iso{25}Mg.
The details of how the \iso{13}C pocket 
forms is still unknown although various mechanisms
have been proposed; see \citet{lugaro04} for a wider discussion.
In nucleosynthesis studies, the extent
of the pocket is usually set as a free parameter with typical values
$\sim\,$1/15$^{\rm th}$ the mass of the He-intershell \citep{gallino98,gm00}.
In intermediate mass stars, the mass of the He-intershell is smaller by about
an order of magnitude compared to lower mass stars and hence the importance
of the \iso{13}C pocket may be lessened \citep{gallino98}.
As done in previous nucleosynthesis studies \citep{gallino98,gm00,lugaro04},
we artificially include in the post-processing calculation a PMZ of constant 
mass $M_{\rm pmz} = 1\times 10^{-4} \Msun$ in the 5$\Msun$, $Z=0.02$ model.
We chose a proton profile in which the number of protons decreases 
exponentially with the mass depth below the base of the convective envelope 
exactly in the same way as described in \citet{lugaro04}.
The mass of the He-intershell decreases with evolution and for the solar
metallicity model, the final He-shell mass is about 0.001, so that our 
PMZ region is at maximum 1/10$^{\rm th}$ of the He intershell.
\par 
The inclusion of the PMZ produces small percentage differences
(less than $\sim$10\%) for most species, with the exception of 
\iso{31}P (28\%) and the particle $g$ (104\%).
Hence there are no changes to the \iso{25,26}Mg yields with
the introduction of a PMZ of $M_{\rm pmz} = 1\times 10^{-4} \Msun$.
This conclusion is consistent with the results obtained by \citet{lugaro99}
for the elements Si and Ti and can be extended to all
intermediate-mass elements. There are two reasons for this. First, 
since elements lighter than iron have very small neutron capture cross 
sections, as much as 3 orders of magnitude smaller as compared to those 
of nuclei heavier than iron, their abundances are not strongly affected 
by the neutron flux in the \iso{13}C pocket.
In fact, the changes in the abundances of intermediate mass elements in 
the \iso{13}C pocket are at maximum of one order of magnitude, while the 
abundances of heavy elements are produced by up to three orders of magnitudes 
of their initial values (see e.g. Table 2 of \citet{lugaro99}). 
Second, the \iso{13}C pocket is engulfed and diluted (of factors between 
1/20$^{\rm th}$ 
to 1/10$^{\rm th}$) by the next growing convective pulse, and is thus mixed with
material already $\alpha$ processed during the previous pulses, and with 
the ashes of the H-burning shell. During this dilution process the signature on 
intermediate mass elements of neutron captures in the \iso{13}C pocket is 
completely lost. As for proton captures occurring in the PMZ region just 
after the mixing of protons has occurred, they are expected to destroy 
\iso{25}Mg only in the upper layer of the PMZ
where the number of protons is higher than 0.5 (see Figure 1 of \citet{gm00}).
Thus they do not produce a strong effect when this region is diluted in
the He intershell, unless the proton profile is very different 
from the one assumed here.
\par
We also experiment with varying the mass-loss rate.
The results for the $Z=0.02$ case with R75 mass loss are shown
in figure~\ref{m5z02r75}. In this figure we observe large 
differences of a few hundred percent for most species.
For the 5$\Msun$, $Z=0.0001$ R75 model we use two different
choices of the \iso{22}Ne $+ \, \alpha$ reaction rates, 
as described in figure~\ref{m5z0001r75}.
The R75 model has fewer TPs than the VW93 case which results
in smaller yields, hence the negative values. 
We see from figure~\ref{m5z0001r75} that changing the
reaction rates has only a small effect on the yields, compared
to the change induced by varying the mass-loss rate. This 
was also well demonstrated by \citet{ventura05b} for intermediate
mass AGB stars of $Z=0.001$. 
The species most affected by the change in mass-loss law 
are \iso{23}Na, \iso{24}Mg and \iso{26}Al, because the abundances 
of these isotopes depend on the duration of the HBB phase.
 
\subsection{Other model uncertainties} 
\par
With regards to the reaction rates, uncertainties remain related to the
proton capture rates for the NeNa and MgAl chains that modify the
abundances of these nuclei because of HBB. Table~\ref{tab:mgyields} 
shows that updating
these proton capture rates from our standard set to those
recommended by NACRE did not produce any major modifications on the Mg
isotopes, since variations are only a few percent.
The variations are small because the reaction rates in the MgAl chain
which affect the Mg isotopic ratios the most are similar between
NACRE and our standard choices, with differences typically
less than 40\%.   A detailed study of
the impact of the uncertainties of proton capture reactions on the results
produced by hot bottom burning on the Ne, Na, Mg and Al isotopes is
underway (Izzard, Lugaro, Karakas \& Iliadis, 2005, in preparation).
\par
We discussed in the introduction that major uncertainties are still present in
the computation of the stellar models, in particular in relation to the
treatment of convection and of convective boundaries.
When our results for the 5$\Msun$, $Z=0.004$ model, which are calculated using
MLT with $\alpha=1.75$, are compared with those produced by a 5$\Msun$, $Z=0.001$
model computed using the ``Full Spectrum of Turbulence'' (FST) prescription
for convective regions by \citet{ventura05a}, we find that the
final abundances of the Mg isotopes are lower in the FST case with respect
to our case by about 2~dex for \iso{24}Mg, 1~dex for \iso{25}Mg and 
0.4~dex for \iso{26}Mg (including the contribution of radioactive 
\iso{26}Al).   In particular, in our case, \iso{25}Mg and \iso{26}Mg 
are produced in similar abundances by the effect of 
the \iso{22}Ne $+ \, \alpha$ reactions and deep TDU, while, 
in the FST case, the production of nuclei of mass A$=$26 (\iso{26}Al and \iso{26}Mg) 
is favored with respect to that of \iso{25}Mg  by the combined effect 
of strong HBB and weak TDU (Ventura P. 2005, private communication). These 
different behaviors should be tested against data from pre-solar grains and, 
using GCE models, against the observations of the Mg isotopes in stars, 
and also the galactic abundance of the radioactive nucleus \iso{26}Al, 
as obtained by the galactic gamma line at 1.809~MeV \citep{diehl04}.
\par 
\citet{herwig04b} implements a diffusive convective overshoot 
scheme at all convective borders in his AGB models 
\citep[see also][]{herwig97}, which results in a considerably 
different stellar structure to our models of the same mass
and metallicity. The most significant differences are hotter TPs
and deeper TDU (where $\lambda \gtrsim 1$), as well as the mixing 
of some CO core material into the flash-driven convective pocket. 
If we compare the surface \iso{25,26}Mg mass fractions after 
the 14$^{\rm th}$ TP, we find that these isotopes are enriched
by up to 5 times more in Herwig's 5$\Msun$, $Z=0.0001$ model
compared to our equivalent case with the new recommended rates.
These differences are much larger than those introduced
by the \iso{22}Ne $+ \, \alpha$ capture rates.
Herwig also reports $X(\iso{26}{\rm Mg})/X(\iso{25}{\rm Mg}) > 1$
after 14 TPs, whereas after the same number of pulses we find
this ratio to be less than unity.
This is because hotter
TPs result in a stronger activation of the 
\iso{22}Ne($\alpha, n$)\iso{25}Mg reaction hence more neutrons 
are released that are captured by \iso{25}Mg to form \iso{26}Mg. 

\section{Discussion \& Conclusions} \label{section:discussion}

\par
The main important result of the present study is that the 
reduction of the uncertainties on the \iso{22}Ne $+ \, \alpha$ 
reaction rates has allowed us to considerably reduce the 
uncertainties coming from these rates on the production of isotopes from 
Mg to P in AGB stars of intermediate mass. The uncertainties on the
Mg yields are now at a level of $\sim 30$\%, much lower than
those obtained when using the NACRE upper or lower limits.
The yields of \iso{25}Mg and \iso{26}Mg are 20\% to 45\% and 
9\% to 16\%, respectively, smaller with the new rates, as compared to 
NACRE. The uncertainties have also been reduced for
species heavier than Mg, where for example, the uncertainty 
on the production of P in solar metallicity models is now at a 
level of 35\%, much lower than the 400\% obtained when using the 
NACRE upper limit. These results are clearly illustrated by 
figures~\ref{yield-compare1} 
and~\ref{yield-compare2} and in table~\ref{tab:mgyields}.
\par
From our analysis it would appear that among the uncertainties 
related to the stellar models, those coming from the treatment of 
convection and of the mass loss are the largest. We described earlier
that these are enormous when compared to the current uncertainties
coming from the \iso{22}Ne $+ \, \alpha$ reaction rates.
Much effort is needed to improve the situation for AGB models,
in particular with respect to convection, by trying to evaluate
and reduce the uncertainties, perhaps by exploiting all the available
observational constraints.
\par
Our new evaluation of the \iso{22}Ne $+ \alpha$ reaction rates will encourage
much future work, as these rates are important for many nucleosynthetic 
processes and sites. The \iso{22}Ne($\alpha, n$)\iso{25}Mg reaction
is an important source of neutrons both during the final evolutionary
stages of massive stars, and during the AGB phase of low to intermediate
mass stars. It is responsible for the production of heavy s-process
elements in these environments \citep{gallino98,rauscher02}. 
The new rate and its uncertainties have to be tested in relation to
this process.  Moreover, the smaller uncertainties of the rates presented
here with respect to those in NACRE, appear to rule out the possibility
that the production of the relatively abundant p-only isotopes of Mo and
Ru could be related to a high value of the \iso{22}Ne($\alpha, n$)\iso{25}Mg 
rates \citep{costa00}.
\par
It is important to study the relative production of \iso{25}Mg and 
\iso{26}Mg,  as we have done in table~\ref{tab:mgyields}, 
because both spectroscopic observations and the analysis of pre-solar grains 
are able to separate these two isotopes. Also the contribution of
radioactive \iso{26}Al to the abundance of \iso{26}Mg has to be 
carefully evaluated. In particular, one pre-solar spinel grain
\citep[][OC2]{zinner05}  
appears to bear the signature of nucleosynthesis in intermediate 
mass AGB stars, with excesses in both  \iso{25}Mg and \iso{26}Mg.
A future application of our present work will be to compare our detailed 
results to the composition of this grain, extending the study to the 
oxygen isotopic ratios in AGB stars and their uncertainties, 
as these are also measured in the grain.



\acknowledgments

\section*{Acknowledgments}

The authors wish to thank the anonymous referee for many helpful comments
that have improved the clarity of the paper.
AIK wishes to acknowledge the Canada Foundation for Innovation (CFI)
and the Nova Scotia Research and Innovation Trust fund (NSRIT) for partly
funding computational resources used for this study. Financial support from
R.G. Deupree's Canada Research Chair (CRC) fund is gratefully 
acknowledged. AIK warmly thanks Maria Lugaro, Brad Gibson and John
Lattanzio for their hospitality, and acknowledges the Institute of
Astronomy (University of Cambridge), the Supercomputing and Astrophysics 
department (Swinburne Institute of Technology) and the CSPA 
(Monash University) for travel support.
We thank Paolo Ventura for sharing unpublished results on the Mg isotopes.



\appendix
\section{Details of the reaction rates used in the reference case}
\label{app:rates}

Most of the 506 reaction rates come from the REACLIB Data Tables (1991 version).
The updates made to the proton, $\alpha$ and neutron capture rates are detailed
in \citet{lugaro04} with one exception. We have since updated the 
\iso{14}N($\alpha,\gamma$)\iso{18}F reaction rate to that given by \citet{gorres00}. 
This reaction rate set is considered our {\em standard} set. Note that in our standard
reaction rate set, we use the \iso{22}Ne $+ \alpha$ rates given by \citet{kaep}
and \citet{drotleff}.

We have also used NACRE rates for many of the reactions involved in the NeNa and MgAl 
chains.  In Table~\ref{tab:pcaptures} we list the reactions that we changed to those
given by NACRE; we also include the reference for that rate used in the standard 
reaction rate set for comparison. CF88 refers to \citet{cf88}.




\clearpage


\begin{table}
\caption{\label{(a,n)} Level parameters for $\alpha$ unbound
states in $^{26}$Mg. }

\begin{tabular}{ccccccc}
\tableline
$E_x$ ($MeV$)&$E_{\alpha}^{cm}$ ($MeV$)&$J^{\pi}$& $\Gamma_{\gamma}$ ($eV$)\tablenotemark{a} & $\Gamma_{n}$
($eV$)\tablenotemark{a} &  $\Gamma_{\alpha}$ ($eV$)\tablenotemark{c} &  $\Gamma_{tot}$ ($eV$ )\tablenotemark{a} \\ \hline
10.693 &  0.078 & 4$^+$ & 3 $^b$ & &
1.7$^{+18.3}_{-1.6}\cdot$10$^{-46}$ & 3\\ \tableline
10.945 &  0.33 & 2$^+$,3$^-$ & 3\tablenotemark{b} & &
6.5$^{+3.9}_{-6.2}\cdot$10$^{-15}$ & 3\\ \tableline
11.112 & 0.497 & 2$^+$ & 1.73 & 2577 &
7.3$^{+14.5}_{-6.3}\cdot$10$^{-11}$ & 2580\\ \tableline
11.153 & 0.538 & 1$^-$ & 4.79 & 14.4 &
4.1$^{+7.6}_{-3.8}\cdot$10$^{-9}$ & 19.2\\ \tableline
11.163 & 0.548 & 2$^+$ & 4.56 & 4640 &
8.4$^{+5.2}_{-3.5}\cdot$10$^{-10}$ & 4650\\ \tableline
11.171 & 0.556 & 2$^+$ & 3\tablenotemark{a,b} & 1.44 &
1.2$^{+0.6}_{-0.6}\cdot$10$^{-9}$ & 20\\ \tableline
11.183 & 0.568 & 1$^-$ & 3\tablenotemark{b} & 0.54 &
7.6$^{+2.6}_{-2.6}\cdot$10$^{-9}$ & 3.54\\ \tableline
11.194 & 0.579 & 2$^+$ & 3\tablenotemark{b} & 0 &
3.1$^{+1.6}_{-1.6}\cdot$10$^{-9}$ & 3\\ \tableline
11.274 & 0.659 & 2$^+$ & 3.24 & 540 &
6.1$^{+2.1}_{-2.1}\cdot$10$^{-8}$ & 543\\ \tableline
11.286 & 0.671 & 1$^-$ & 0.79 & 1256 &
3.4$^{+1.2}_{-1.2}\cdot$10$^{-7}$ & 1257\\ \tableline
11.310 & 0.695 & 1$^-$ & 3\tablenotemark{b} & 1.12 &
5.8$^{+1.5}_{-1.5}\cdot$10$^{-6}$ & 4.1\\ \tableline
11.326 & 0.711 & 1$^-$ & 3\tablenotemark{b} & 0.6 &
9.4$^{+3.2}_{-3.2}\cdot$10$^{-6}$ & 3.6\\ \tableline
11.328 & 0.713 & 1$^-$ & 3.6 & 424 &
3.9$^{+1.3}_{-1.3}\cdot$10$^{-5}$ & 428\\ \tableline
\end{tabular}
\tablenotetext{a}{reference \citet{koehler}}
\tablenotetext{b}{estimated average value \citep{koehler}}
\tablenotetext{c}{reference \citet{giesen}}
\end{table}

\clearpage

\begin{table}
\caption{Recommended resonance strengths as well as their upper and lower limits
for the $^{22}$Ne($\alpha,\gamma$)$^{26}$Mg reaction derived from the
parameters listed in table \ref{(a,n)}\label{ogres1}} 
\begin{tabular}{cccccc}
\tableline
$E_x$ ($MeV$)&$E_{\alpha}^{cm}$ ($MeV$)&$J^{\pi}$&
$\omega\gamma_{(\alpha,\gamma)}$ ($eV$)&
$\omega\gamma_{(\alpha,\gamma)}^{ll}$ ($eV$)&
$\omega\gamma_{(\alpha,\gamma)}^{ul}$ ($eV$)\\ \tableline
10.693 &  0.078 & 4$^+$       & 1.6$\cdot$10$^{-45}$& 1.2$\cdot$10$^{-46} $ &3.6$\cdot$10$^{-44} $\\ 
10.945 &  0.330 & 2$^+$,3$^-$ & 2.8$\cdot$10$^{-15}$& 1.8$\cdot$10$^{-15} $ &1.0$\cdot$10$^{-13} $\\ 
11.112 &  0.497 & 2$^+$       & 2.4$\cdot$10$^{-13}$& 3.4$\cdot$10$^{-14} $ &7.3$\cdot$10$^{-13} $\\ 
11.153 &  0.538 & 1$^-$       & 3.0$\cdot$10$^{-09}$& 2.1$\cdot$10$^{-10} $ &8.7$\cdot$10$^{-09} $\\ 
11.163 &  0.548 & 2$^+$       & 4.1$\cdot$10$^{-12}$& 2.4$\cdot$10$^{-12} $ &6.7$\cdot$10$^{-12} $\\ 
11.171 &  0.556 & 2$^+$       & 8.9$\cdot$10$^{-10}$& 4.5$\cdot$10$^{-10} $ &1.3$\cdot$10$^{-09} $\\ 
11.183 &  0.568 & 1$^-$       & 1.9$\cdot$10$^{-08}$& 1.3$\cdot$10$^{-08} $ &2.4$\cdot$10$^{-08} $\\ 
11.194 &  0.579 & 2$^+$       & 1.6$\cdot$10$^{-08}$& 7.5$\cdot$10$^{-09} $ &2.4$\cdot$10$^{-08} $\\ 
11.274 &  0.659 & 2$^+$       & 1.8$\cdot$10$^{-09}$& 1.2$\cdot$10$^{-09} $ &2.5$\cdot$10$^{-09} $\\ 
11.286 &  0.671 & 1$^-$       & 6.4$\cdot$10$^{-10}$& 4.2$\cdot$10$^{-10} $ &8.7$\cdot$10$^{-10} $\\ 
11.310 &  0.695 & 1$^-$       & 3.6$\cdot$10$^{-05}$& 3.2$\cdot$10$^{-05} $ &4.0$\cdot$10$^{-05} $\\ 
11.326 &  0.711 & 1$^-$       & 2.3$\cdot$10$^{-05}$& 1.6$\cdot$10$^{-05} $ &3.2$\cdot$10$^{-05} $\\ 
11.328 &  0.713 & 1$^-$       & 9.9$\cdot$10$^{-07}$& 6.8$\cdot$10$^{-07} $ &1.3$\cdot$10$^{-06} $\\ 
\tableline
\end{tabular}
\end{table}

\clearpage

\begin{table}
\caption{Recommended resonance strengths as well as their upper and lower limits
for the $^{22}$Ne($\alpha,n$)$^{25}$Mg reaction derived from the
parameters listed in Table~\ref{(a,n)}. \label{ogres2}}
\begin{tabular}{cccccc}
\tableline
$E_x$ ($MeV$)&$E_{\alpha}^{cm}$ ($MeV$)&$J^{\pi}$&
$\omega\gamma_{(\alpha,n)}$ ($eV$)&
$\omega\gamma_{(\alpha,n)}^{ll}$ ($eV$)&
$\omega\gamma_{(\alpha,n)}^{ul}$ ($eV$) \\ \tableline
11.112 &  0.497 & 2$^+$ &  3.6$\cdot$10$^{-10} $ &  5.0$\cdot$10$^{-11} $ &  1.1$\cdot$10$^{-09} $ \\ 
11.153 &  0.538 & 1$^-$ &  9.2$\cdot$10$^{-09} $ &  6.3$\cdot$10$^{-10} $ &  2.6$\cdot$10$^{-08} $ \\ 
11.163 &  0.548 & 2$^+$ &  4.2$\cdot$10$^{-09} $ &  2.4$\cdot$10$^{-09} $ &  6.8$\cdot$10$^{-09} $ \\ 
11.171 &  0.556 & 2$^+$ &  4.3$\cdot$10$^{-10} $ &  2.2$\cdot$10$^{-10} $ &  6.5$\cdot$10$^{-10} $ \\ 
11.183 &  0.568 & 1$^-$ &  3.5$\cdot$10$^{-09} $ &  2.3$\cdot$10$^{-09} $ &  4.6$\cdot$10$^{-09} $ \\ 
11.274 &  0.659 & 2$^+$ &  3.0$\cdot$10$^{-07} $ &  1.0$\cdot$10$^{-07} $ &  4.1$\cdot$10$^{-07} $ \\ 
11.286 &  0.671 & 1$^-$ &  1.0$\cdot$10$^{-06} $ &  6.6$\cdot$10$^{-07} $ &  1.4$\cdot$10$^{-06} $ \\ 
11.310 &  0.695 & 1$^-$ &  4.7$\cdot$10$^{-06} $ &  3.5$\cdot$10$^{-06} $ &  6.0$\cdot$10$^{-06} $ \\ 
11.326 &  0.711 & 1$^-$ &  4.7$\cdot$10$^{-06} $ &  3.1$\cdot$10$^{-06} $ &  6.5$\cdot$10$^{-06} $ \\ 
11.328 &  0.713 & 1$^-$ &  1.18$\cdot$10$^{-04} $&  1.03$\cdot$10$^{-04} $&  1.33$\cdot$10$^{-04} $ \\ \tableline
\end{tabular}
\end{table}

\clearpage

\begin{table}
\caption{Structural properties of the 5$\Msun$ AGB models, see the text for details.  \label{table:agbmodels}}
\begin{tabular}{rrcccccccccc}
\tableline
$Z$ & $\dot{M}$ & $M_{\rm c}(1)$ & $M_{\rm env}^{\rm HBB}$ & $M_{\rm env}^{\rm f}$ &
No. & No. & $M_{\rm dred}^{\rm tot}$ & $T_{\rm BCE}^{\rm max}$ & 
$T_{\rm He}^{\rm max}$ & $\lambda_{\rm max}$ \\
  &  & ($\Msun$) & ($\Msun$) & ($\Msun$) & TP & TDU  & ($\Msun$) & ($10^{6}$K) & ($10^{6}$K) &  \\
\tableline\tableline
0.02   & VW93 & 0.861 & 2.564 & 1.499 & 24 & 22 & 5.027($-2$) & 64 & 352 & 0.961  \\
0.02   & R75  & 0.861 & 2.560 & 1.802 & 38 & 35 & 1.047($-1$) & 57 & 368 & 0.977 \\
0.008  & VW93 & 0.870 & 1.857 & 1.387 & 59 & 56 & 1.745($-1$) & 81 & 366 & 0.952  \\
0.004  & VW93 & 0.888 & 1.560 & 0.944 & 83 & 80 & 2.250($-1$) & 85 & 379 & 0.959  \\
10$^{-4}$ & VW93 & 0.910 & 1.056 & 0.572 & 137 & 134 & 3.133($-1$) & 92 & 380 & 0.980 \\ 
10$^{-4}$ & R75  & 0.909 & 1.100 & 0.335 & 70  & 67  & 1.569($-1$) & 90 & 383 & 0.950  \\
\tableline
\end{tabular}
\end{table}

\clearpage

\begin{table}
\caption{Some surface abundance results for the 5$\Msun$ models including the
final C$+$N$+$O and \iso{12}C/\iso{13}C ratio, and C/O ratio and the Mg isotope ratio 
(\iso{24}Mg:\iso{25}Mg:\iso{26}Mg) at the beginning and end of the TP-AGB phase (using
the new recommended rates).
\label{tab:agb2}}
\begin{tabular}{rrcccccc}
\tableline
$Z$ & $\dot{M}$ & CNO$_{f}$/ & \iso{12}C/\iso{13}C$_{\rm f}$ & C/O$_{\rm agb}$ & C/O$_{\rm f}$ & 
Mg Ratio$_{\rm agb}$ & Mg Ratio$_{\rm f}$ \\
    &           & CNO$_{0}$  &      &            &               &                      &    \\ 
\tableline\tableline
0.02   & VW93 & 1.341 & 7.828 & 0.291 & 0.766 & 79.0:09.6:11.4 & 72.2:12.4:15.4 \\
0.02   & R75  & 1.814 & 15.57 & 0.291 & 1.705 & 79.0:09.6:11.4 & 56.1:18.0:25.9  \\
0.008  & VW93 & 2.886 & 7.383 & 0.239 & 0.660 & 82.8:07.6:09.6 & 46.1:19.6:34.2  \\
0.004  & VW93 & 5.106 & 10.77 & 0.204 & 2.102 & 85.3:06.4:8.33 & 31.7:21.6:46.7 \\
10$^{-4}$ & VW93 & 431.1 & 9.223 & 0.209 & 9.400 & 82.2:08.2:09.6 & 11.7:17.5:70.7 \\
10$^{-4}$ & R75  & 339.4 & 12.15 & 0.205 & 18.60 & 82.2:08.2:09.6 & 04.5:16.7:78.7  \\
\tableline
\end{tabular}
\end{table}

\clearpage

\begin{deluxetable}{ccrrrrr}
\tablecolumns{7}
\tablewidth{0pc}
\tablecaption{\label{tab:mgyields} 
In the first row we show the yields of  \iso{25}Mg (in roman font) and \iso{26}Mg 
(in italics) in solar masses for different stellar models ($M (\Msun)$, $Z$) 
that use the new recommended rates for the \iso{22}Ne $+ \alpha$ reactions.
In other rows, we show the percentage difference between models computed using
other estimates of the \iso{22}Ne $+ \alpha$ reactions and the yields from
row 1 (see definition in the text).} 

\tablehead{\colhead{$^{22}$Ne($\alpha,\gamma)^{26}$Mg} &
\colhead{$^{22}$Ne($\alpha, n$)$^{25}$Mg} & \colhead{(5,0.0001)} &
\colhead{(5,0.004)} & \colhead{(5,0.008)} & \colhead{(5,0.02)} }

\startdata

recommended & recommended & 5.80($-4$) & 4.84($-4$) & 3.49($-4$) & 6.49($-5$) \\
 & & {\it 2.28($-3$)} & {\it 1.11($-3$)} & {\it 6.84($-4$)} & {\it 1.17($-4$)} \\ \hline
standard$^{a}$ & standard$^{a}$ & $-$3.42 & 29.6 & 36.6 & 45.6 \\
 & & {\it 9.05} & {\it 7.50} & {\it 3.51} & {\it $-$5.95} \\ \hline
K94$^{b}$ & K94$^{b}$ & 22.5 & 31.5 & 38.2 & 52.6  \\
 & & {\it 9.80} & {\it 8.64} & {\it 5.68} & {\it $-$2.99} \\ \hline
NACRE$^{c}$ & NACRE$^{c}$ & 20.4 & 27.9 & 34.6 & 44.8  \\
 & & {\it 15.3} & {\it 16.0} & {\it 15.9} & {\it 8.82} \\ \hline
upper limit & upper limit & 3.34 & 8.94 & 11.5 & 24.7  \\
 & & {\it 13.8} & {\it 17.0} & {\it 18.9} & {\it 26.7} \\  \hline
lower limit & lower limit & $-$8.93 & $-$7.95 & $-$9.98 & $-$14.7 \\
 & & {\it $-$14.2} & {\it $-$16.2} & {\it $-$17.5} & {\it $-$16.6} \\  \hline
lower limit & upper limit & 5.10 & 11.5 & 13.5 & 20.7 \\
 & & {\it 0.46} & {\it 0.21} & {\it $-$1.46} & {\it $-$4.13} \\ \hline
upper limit & lower limit & $-$9.95 & $-$8.94 & $-$11.2 & $-$16.5 \\
 & & {\it 2.47} & {\it 3.82} & {\it 4.96} & {\it 7.82} \\

\enddata
\vskip 0.5 cm

{\footnotesize $^a$Standard Case: K\"{a}ppeler et al. (1994, K94) for the \iso{22}Ne $+ \alpha$ reactions;
reaction rates for the NeNa and MgAl chains described in the Appendix}\\
{\footnotesize $^b$K94: K94 for the \iso{22}Ne $+ \alpha$ reactions; NACRE recommended rates for the
NeNa and MgAl chains}\\
{\footnotesize $^c$NACRE: NACRE recommended rates for the \iso{22}Ne $+ \alpha$ reactions and the
NeNa and MgAl chains}\\

\end{deluxetable}

\clearpage

\begin{deluxetable}{lc}
\tablecolumns{2}
\tablewidth{0pc}
\tablecaption{\label{tab:pcaptures}
Proton capture rates changed to those given by NACRE}
\tablehead{\colhead{reaction} & \colhead{reference}}
\startdata

~\iso{19}F($p, {\gamma}$)\iso{20}Ne & CF88 \\

~\iso{20}Ne($p, {\gamma}$)\iso{21}Ne & CF88  \\

~\iso{21}Ne($p, {\gamma}$)\iso{22}Na &  \citet{eleid95} \\

~\iso{22}Ne($p, {\gamma}$)\iso{23}Na$^{a,b}$ & \citet{eleid95}  \\

~\iso{23}Na($p, {\gamma}$)\iso{24}Mg &  \citet{eleid95}  \\

~\iso{23}Na($p, {\alpha}$)\iso{20}Ne  &  \citet{eleid95} \\ 

~\iso{24}Mg($p, {\gamma}$)\iso{25}Al  &  \citet{powell99}  \\

~\iso{25}Mg($p, {\gamma}$)\iso{26}Al$^{g/i}$ & \citet{iliadis96,iliadis90} \\ 

$^{26}$Mg($p, {\gamma}$)$^{27}$Al & \citet{iliadis90} \\

~\iso{26}Al$^{g}$($p, \gamma$)\iso{27}Si   &  \citet{champagne93,vogelaar96} \\

~\iso{26}Al$^{i}$($p, \gamma$)\iso{27}Si   &  CF88  \\

~\iso{27}Al(${p, \gamma}$)\iso{28}Si  &   \citet{iliadis90,timmermann88}  \\

~\iso{27}Al($p, {\alpha}$)\iso{24}Mg  &   \citet{timmermann88,champagne88}  \\

\enddata
 
\flushleft{$^{a}$but beware the error in the NACRE analytical fit
for $0.15 \ge T_{9} \ge 2.0$ compared to the tabulated rate.}\\

\flushleft{$^{b}$ The rate given by \citet{eleid95} contains a typographical error
for the first term.}\\

\end{deluxetable}


\clearpage

\centerline{\bf FIGURE CAPTIONS}

\figcaption[fig1]{\label{fig:mg-previous} The stellar yields of (a) \iso{25}Mg and 
(b) \iso{26}Mg (in $\Msun$) as a function of mass and metallicity, from \citet{kara03}. 
Also included are the yields from the $Z=0.0001$ AGB models. The reaction rates used for
these models are our {\em standard} choice as outlined in the text.
We define the stellar yield to be $y_{k} = \int_{\tau} [X(k) - X_{0}(k)] (dM/dt) dt$,
where $\tau$ is the stellar lifetime, $X(k)$ is the current mass fraction, 
$X_{0}(k)$ the initial mass fraction, and $dM/dt$ is current mass-loss rate.}

\figcaption[fig2]{\label{22ne+a} Reaction rates for $^{22}$Ne($\alpha,n$)$^{25}$Mg and
$^{22}$Ne($\alpha,\gamma$)$^{26}$Mg as a function of temperature.  The solid lines represent 
the uncertainty range in the reaction rate as discussed in the text.}

\figcaption[fig3]{\label{ratio} Ratio of the upper (lower) limits of the 
$^{22}$Ne($\alpha,n$)$^{25}$Mg and $^{22}$Ne($\alpha,\gamma$)$^{26}$Mg reaction 
rates and the recommended values as a function of temperature.}

\figcaption[fig4]{\label{anag} Ratio of the reaction rates of $^{22}$Ne($\alpha,n$)$^{25}$Mg
and $^{22}$Ne($\alpha,\gamma$)$^{26}$Mg as a function of temperature.
Shown are the ratio between the recommended values and the
ratio between the upper limit of $^{22}$Ne($\alpha,n$)$^{25}$Mg
and the lower limit of $^{22}$Ne($\alpha,\gamma$)$^{26}$Mg as upper
limit for the overall uncertainty ratio as well as the ratio between the lower limit of
$^{22}$Ne($\alpha,n$)$^{25}$Mg and the upper limit of $^{22}$Ne($\alpha,\gamma$)$^{26}$Mg
as a lower limit for the ratio.}

\figcaption[fig5]{\label{m5z02} The evolution of the \iso{25}Mg (solid line) and \iso{26}Mg
(dashed line) abundances at the surface of the 5$\Msun$, $Z=0.02$ VW93 model for four 
different choices of the \iso{22}Ne $+ \, \alpha$ reaction rates. }

\figcaption[fig6]{\label{mg-surf} The surface abundance (in $\log$ mole fraction) of the 
Mg (\iso{24}Mg dotted line, \iso{25,26}Mg as in Fig.~\ref{m5z02}) isotopes as a function of 
time during the TP-AGB for models of different metallicity, using the new recommended rates.}

\figcaption[fig7]{\label{yield-compare1} Percentage difference between stellar
yields computed with the new recommended rates (the reference) and the new
upper and lower limits and the NACRE recommended rates for the \iso{22}Ne 
$+ \, \alpha$ reactions. We show results for the 5$\Msun$, $Z=0.02$ (a) and 
for $Z=0.0001$ model (b).}

\figcaption[fig8]{\label{yield-compare2} Same as figure~\ref{yield-compare1}, but 
now the reference yields were computed with the  NACRE \iso{22}Ne $+ \alpha$ 
recommended rates and we show percentage differences for models using the NACRE
upper and lower limits. We also show results where the \iso{25}Mg($n, \gamma$)\iso{26}Mg 
reaction rate was varied by a factor of two each way.}

\figcaption[fig9]{\label{m5z02r75} The percentage difference between the
yields from the 5$\Msun$, $Z=0.02$ model with R75 mass loss.}

\figcaption[fig10]{\label{m5z0001r75}  The percentage difference between the
yields from the 5$\Msun$, $Z=0.0001$ model with VW93 mass loss and two models
with R75 mass loss.}

\clearpage

\begin{figure}
\plottwo{f1a}{f1b}
\end{figure}

\clearpage

\begin{figure}
\plotone{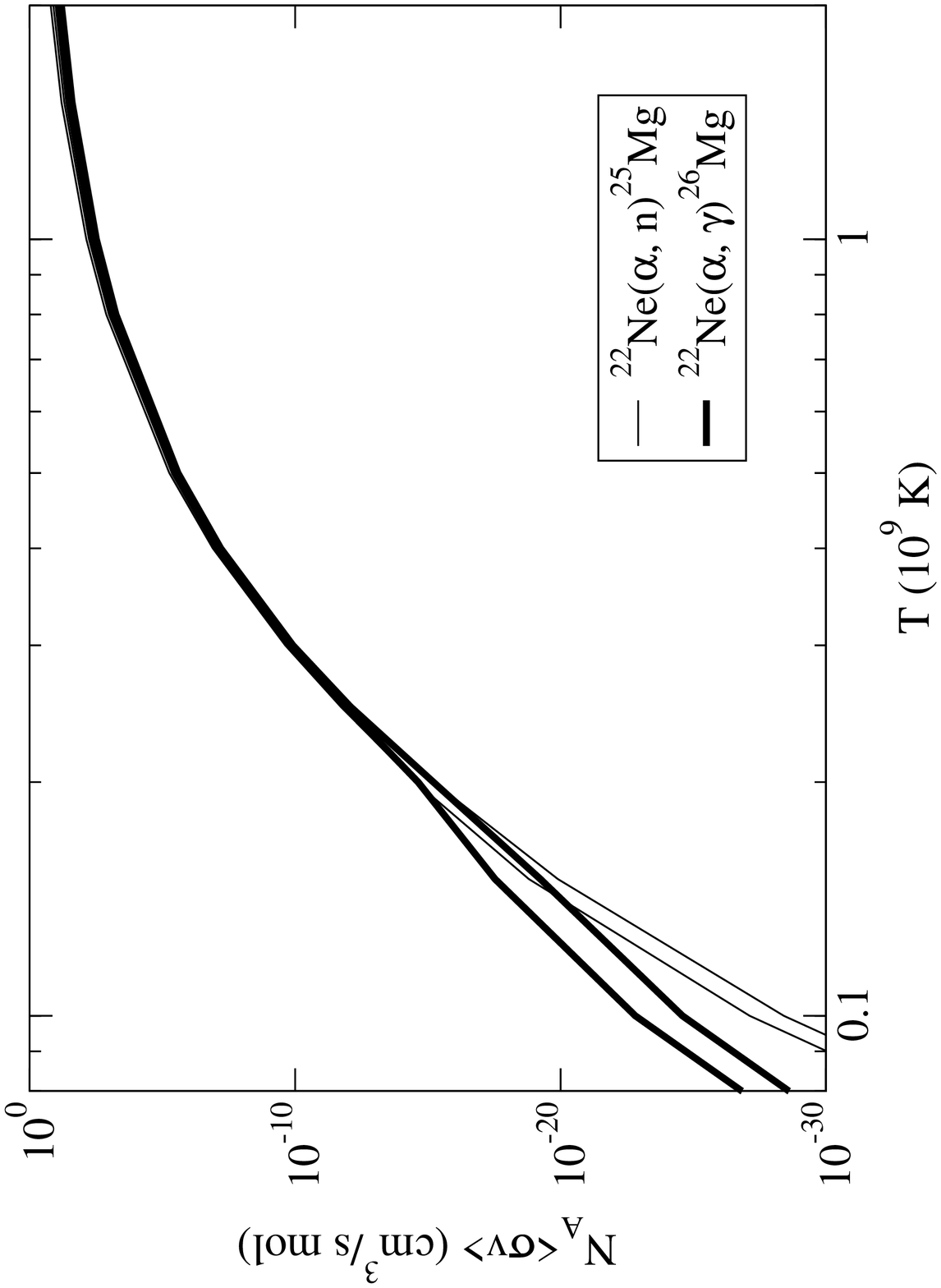}
\end{figure}

\clearpage

\begin{figure}
\plotone{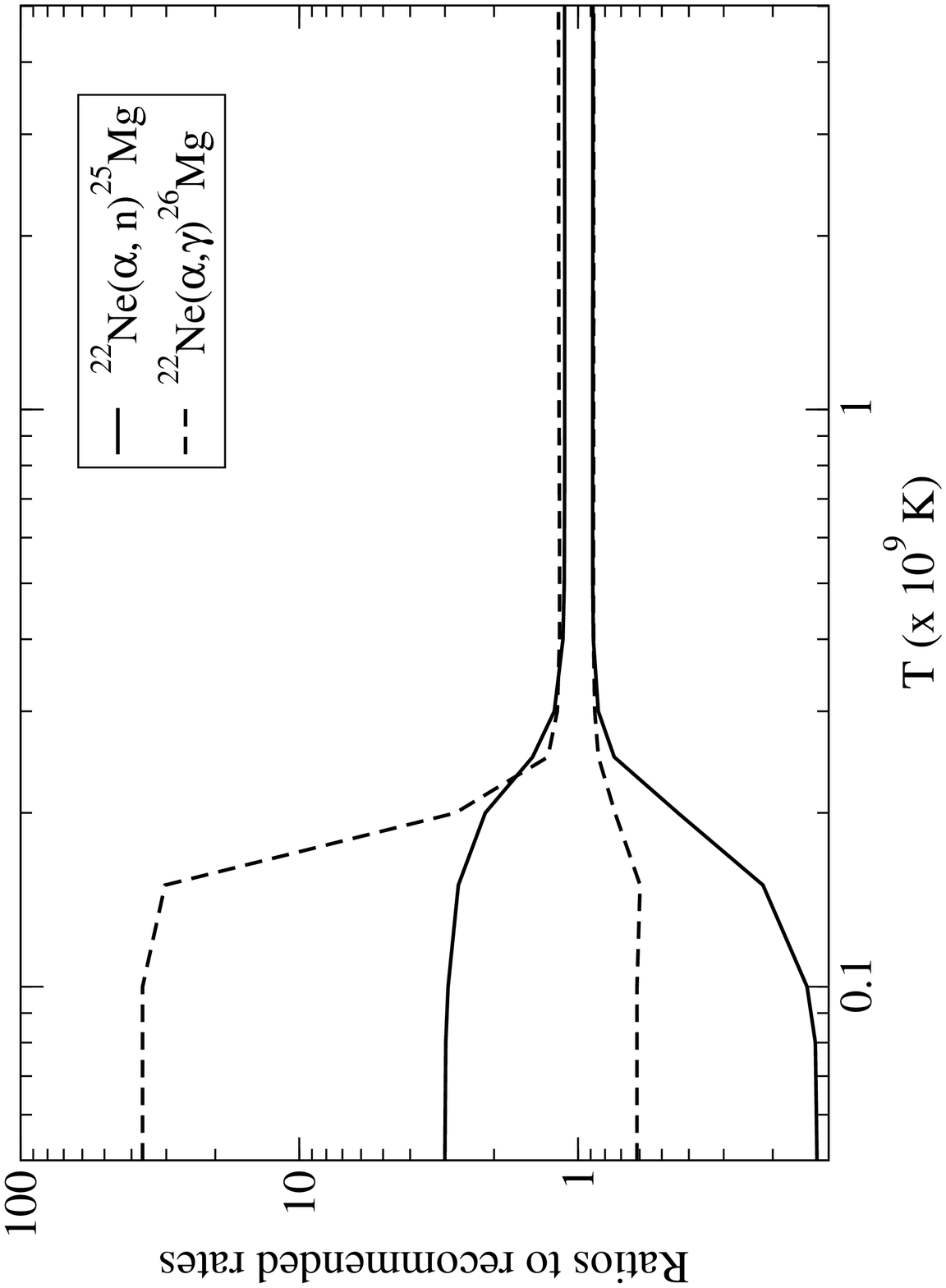}
\end{figure}
 
\begin{figure}
\plotone{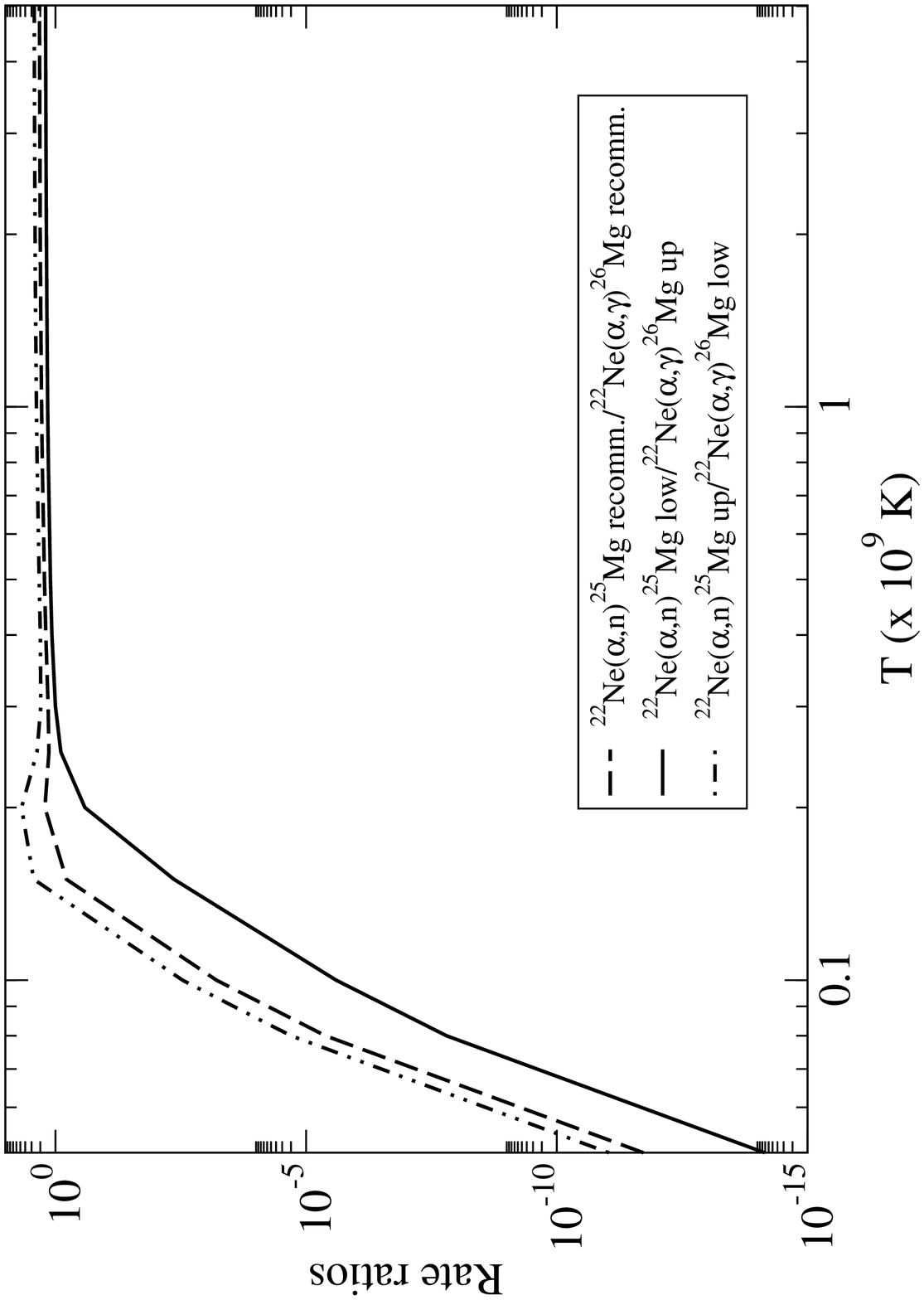}
\end{figure}

\clearpage

\begin{figure}
{\plottwo{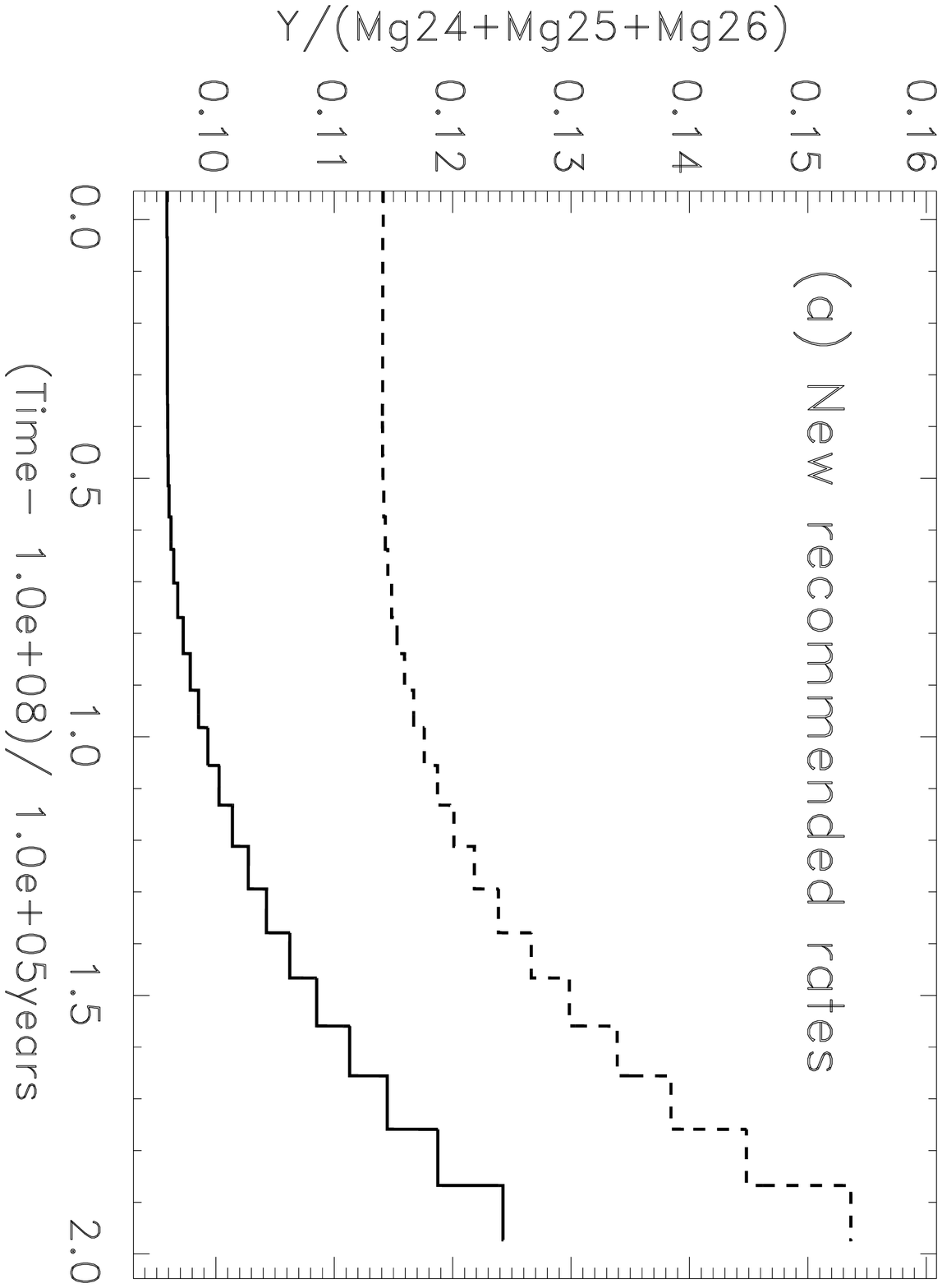}{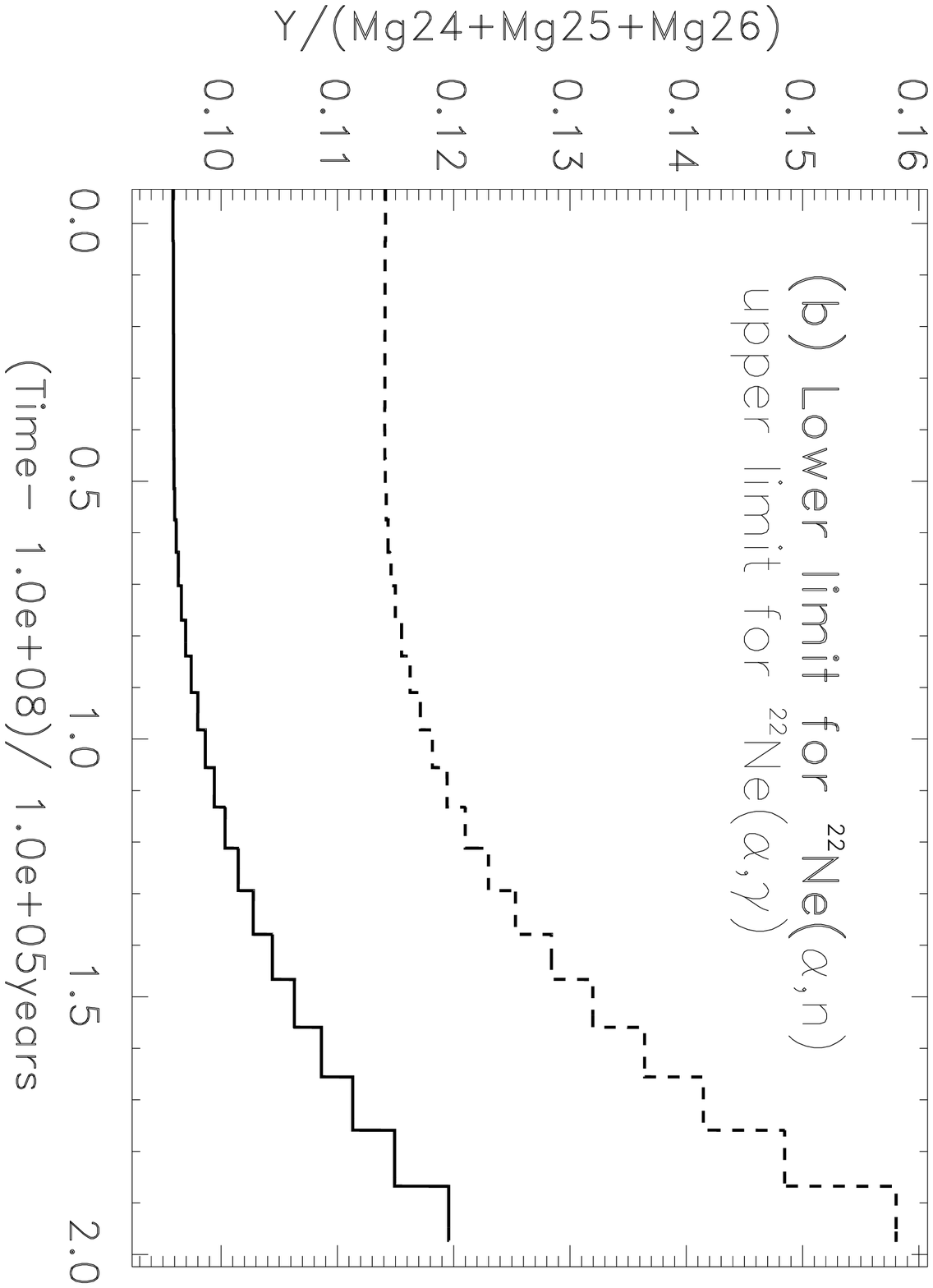}}
{\plottwo{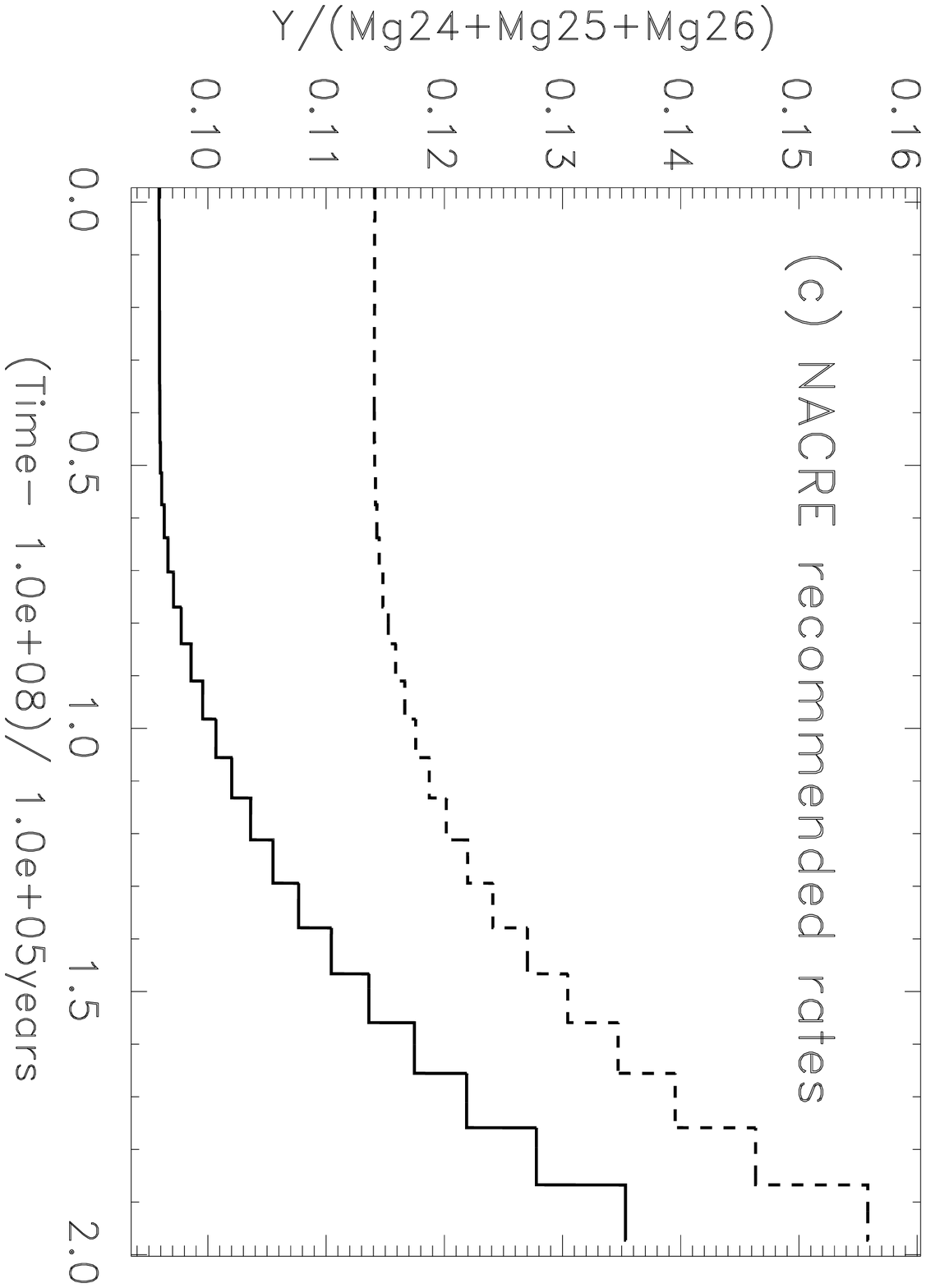}{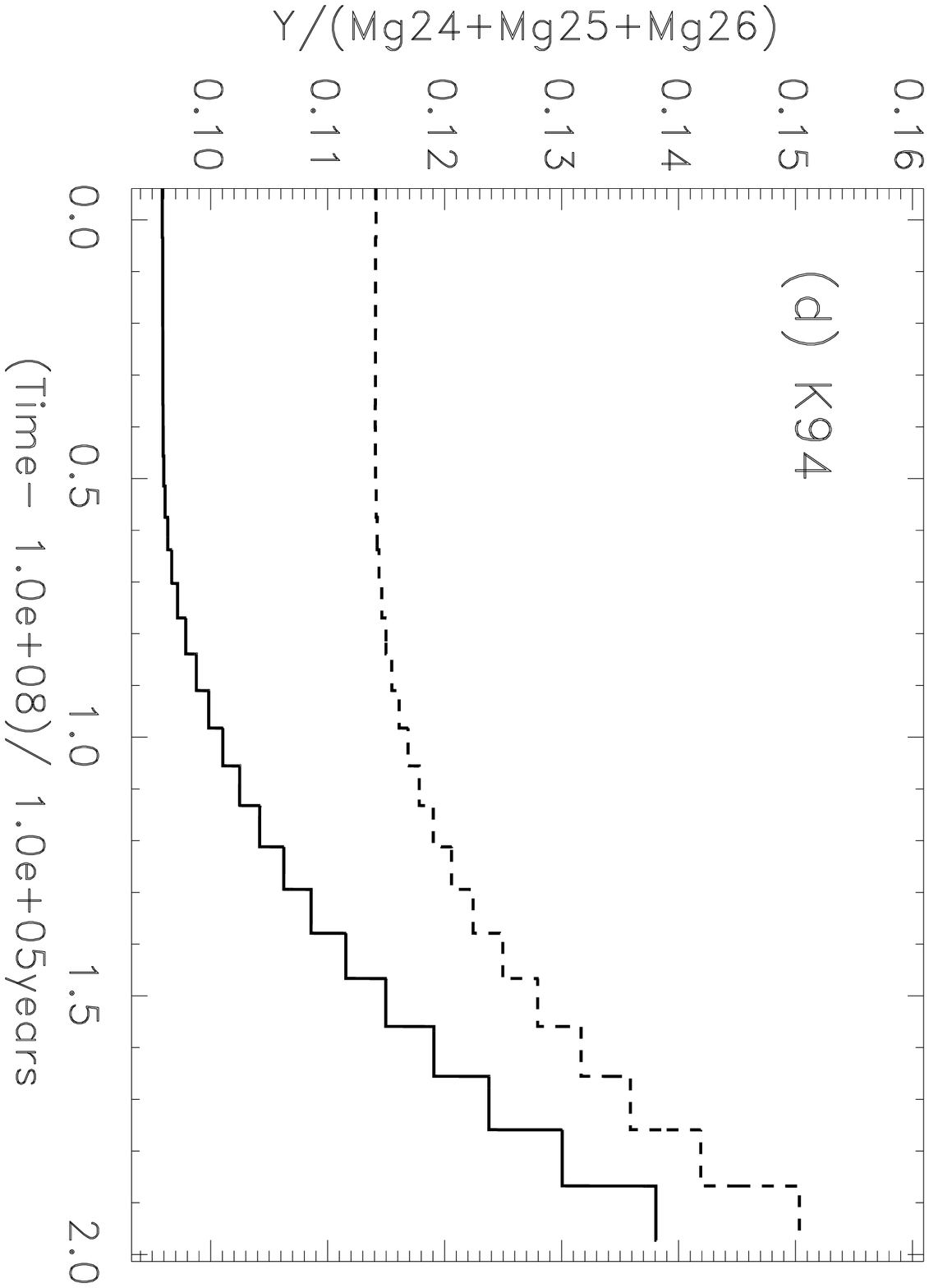}}
\end{figure}

\clearpage

\begin{figure}
\begin{tabular}{c}
\includegraphics[angle=90,width=0.6\textwidth]{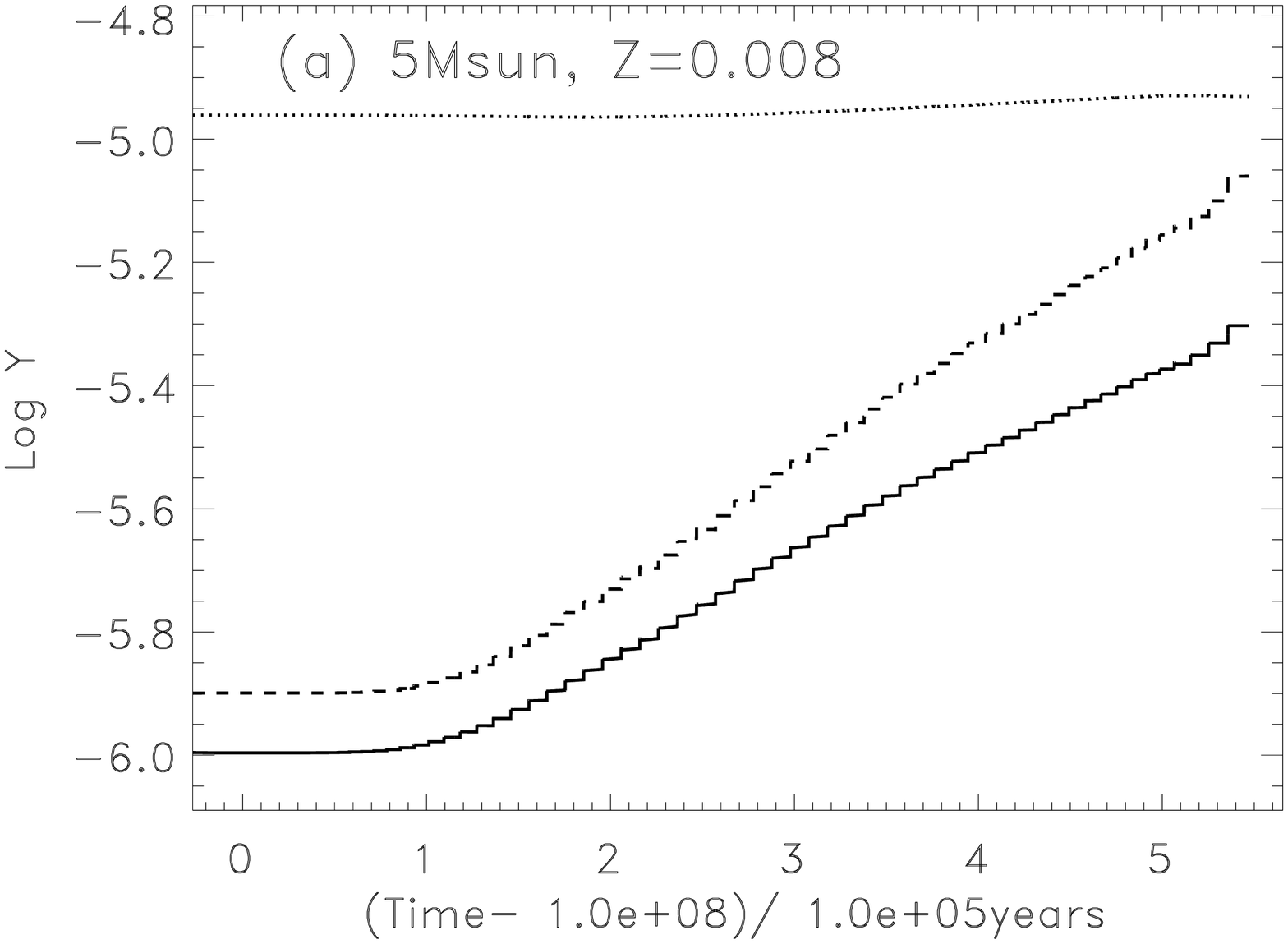} \\
\includegraphics[angle=90,width=0.6\textwidth]{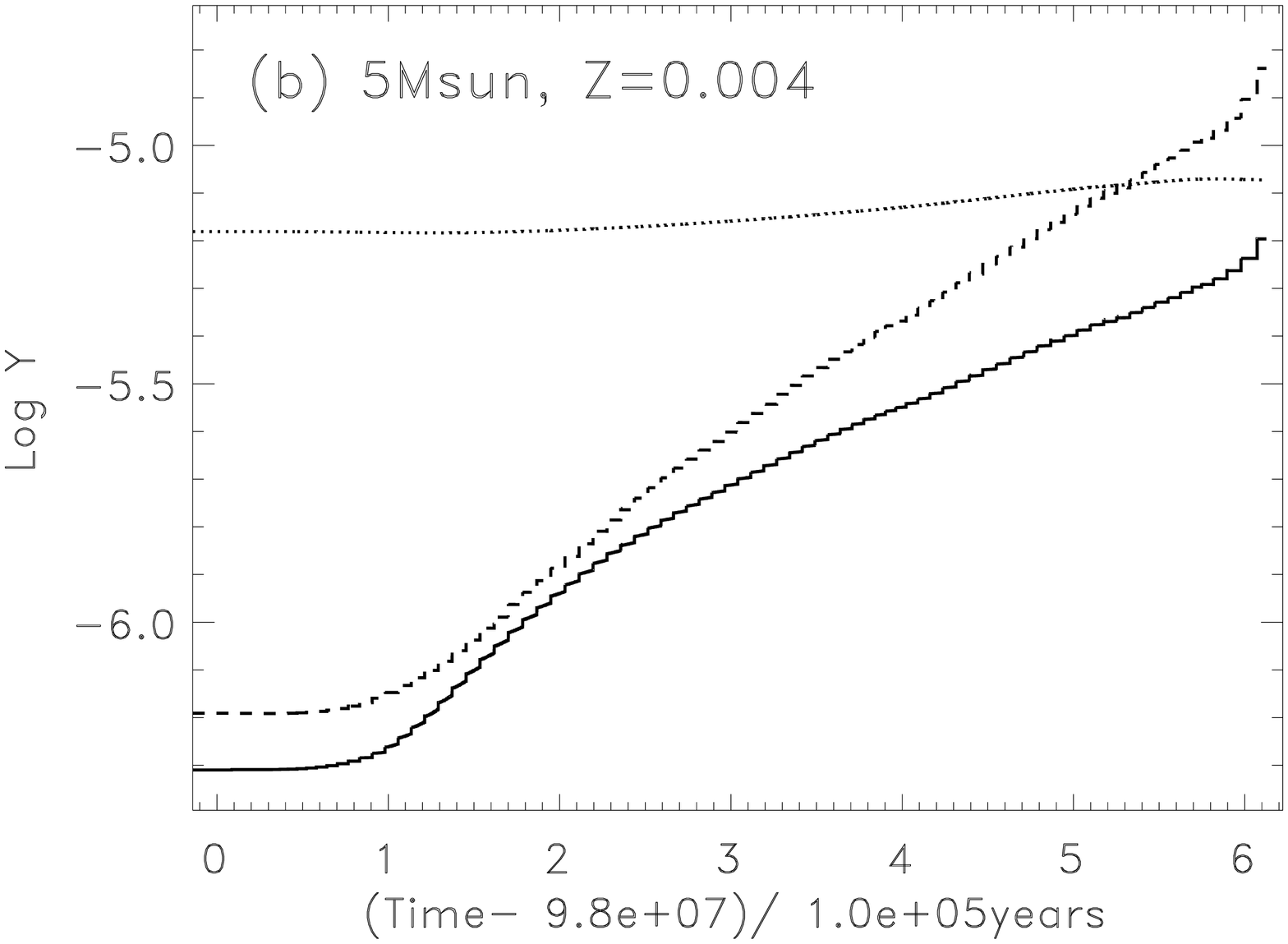} \\
\includegraphics[angle=90,width=0.6\textwidth]{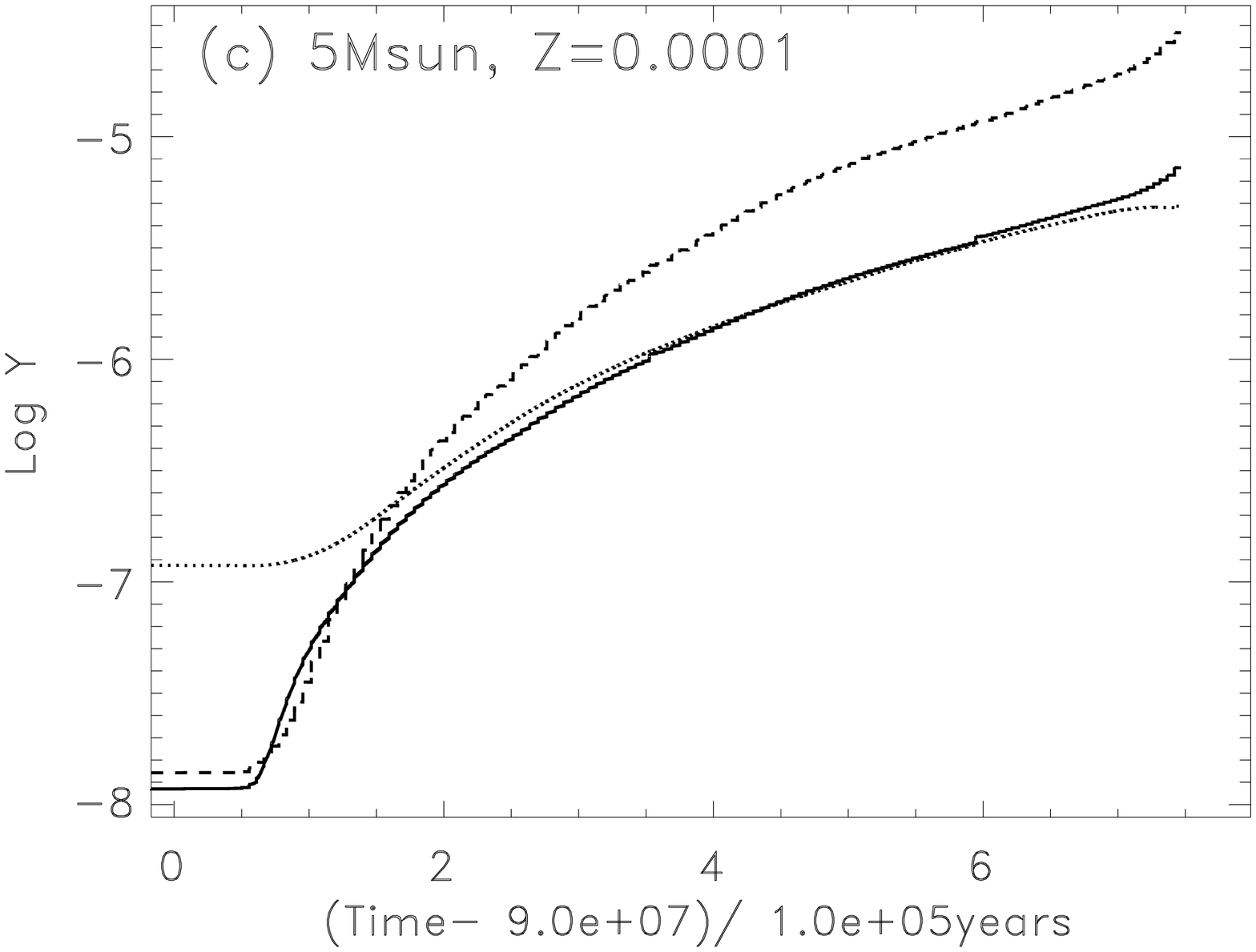} \\
\end{tabular}
\end{figure}
\clearpage

\begin{figure}
\plottwo{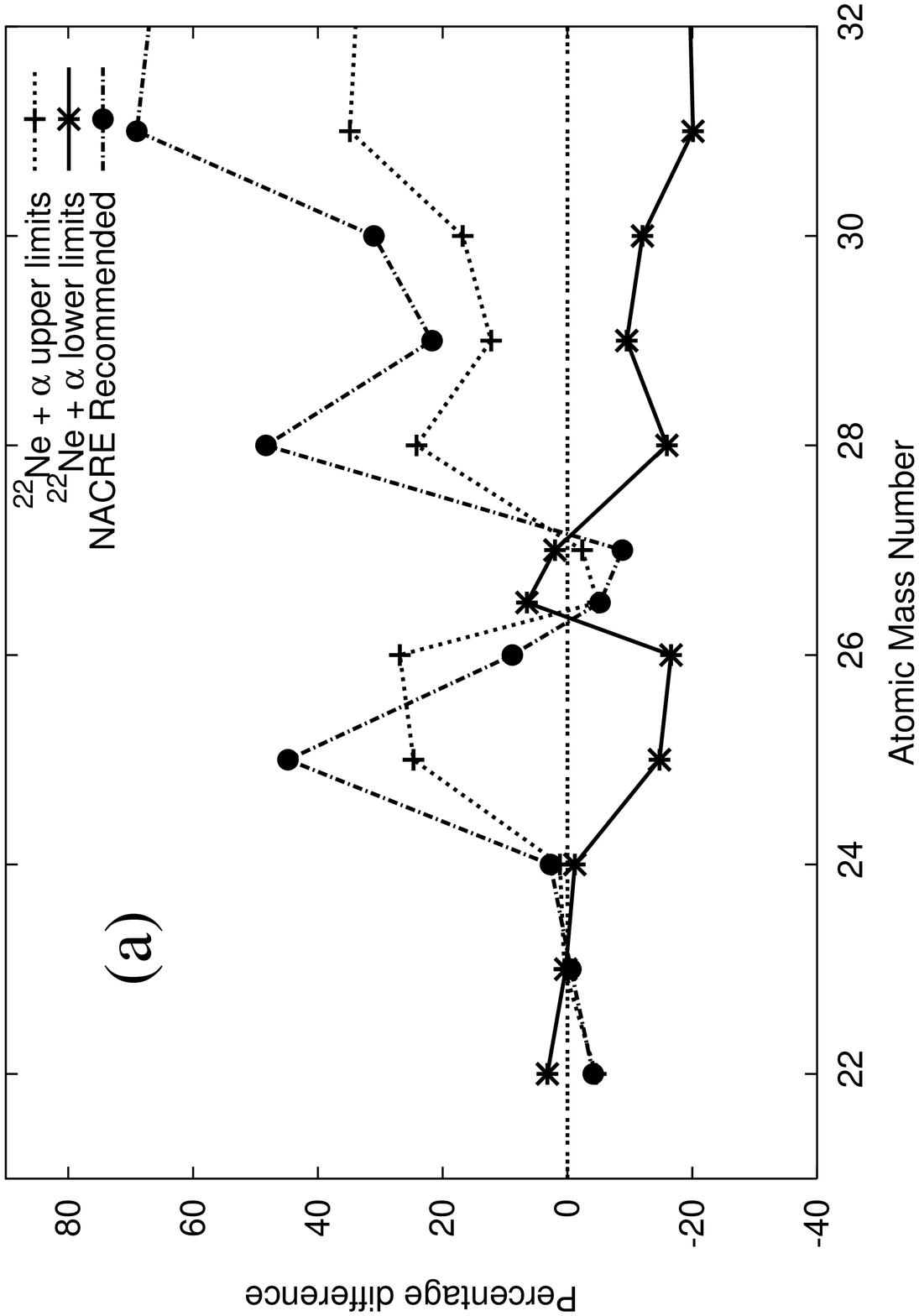}{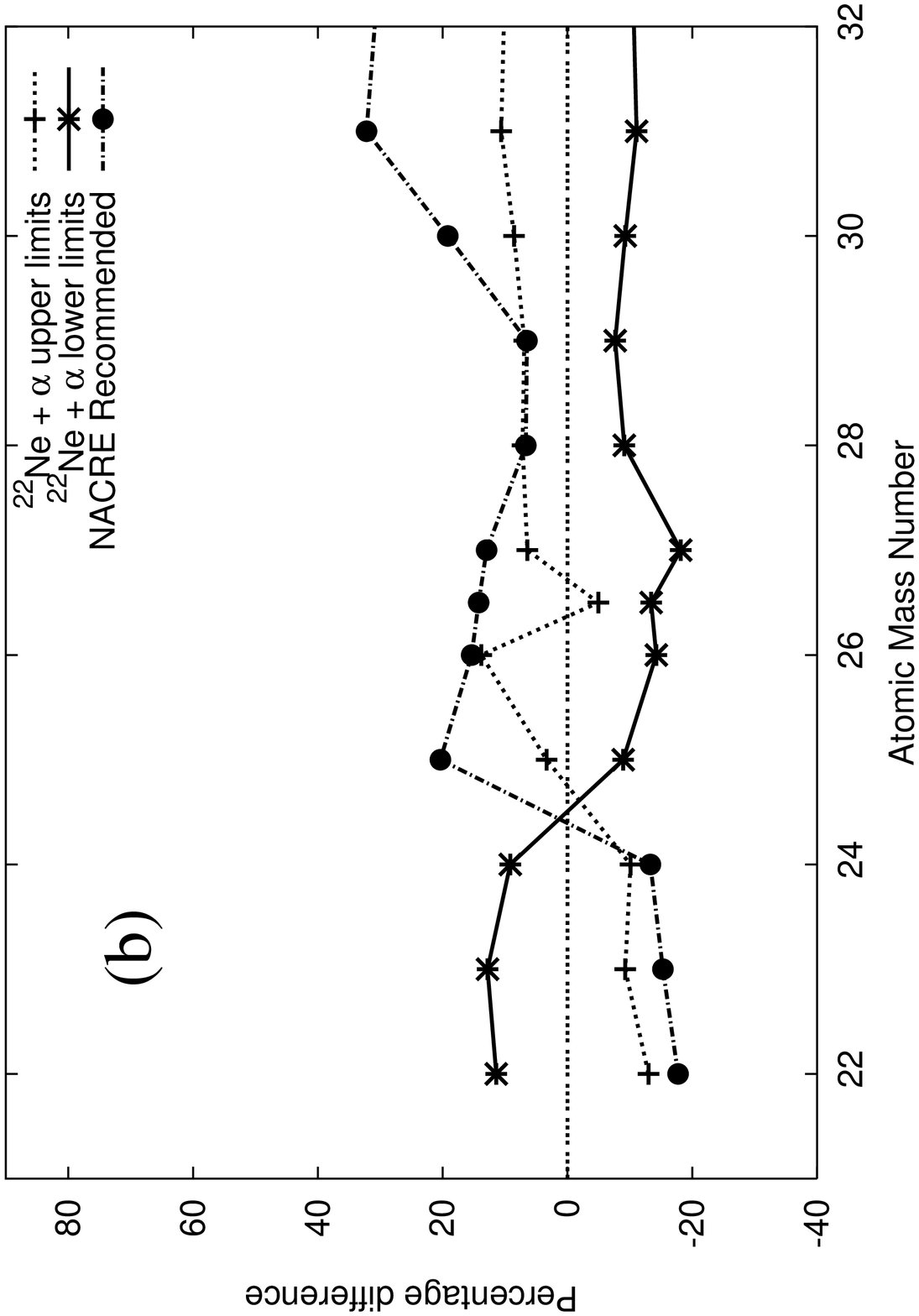}
\end{figure}

\clearpage

\begin{figure}
\plottwo{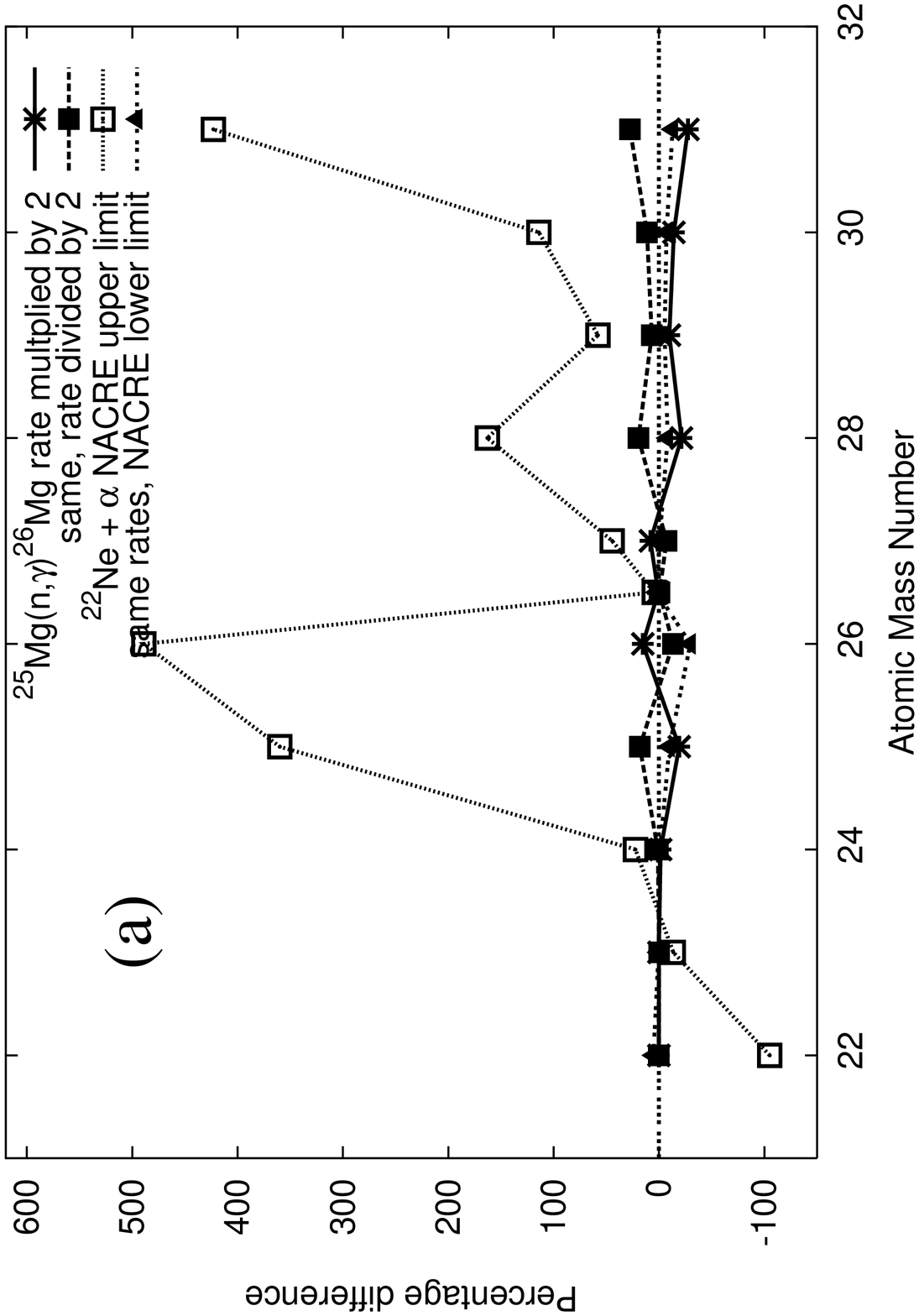}{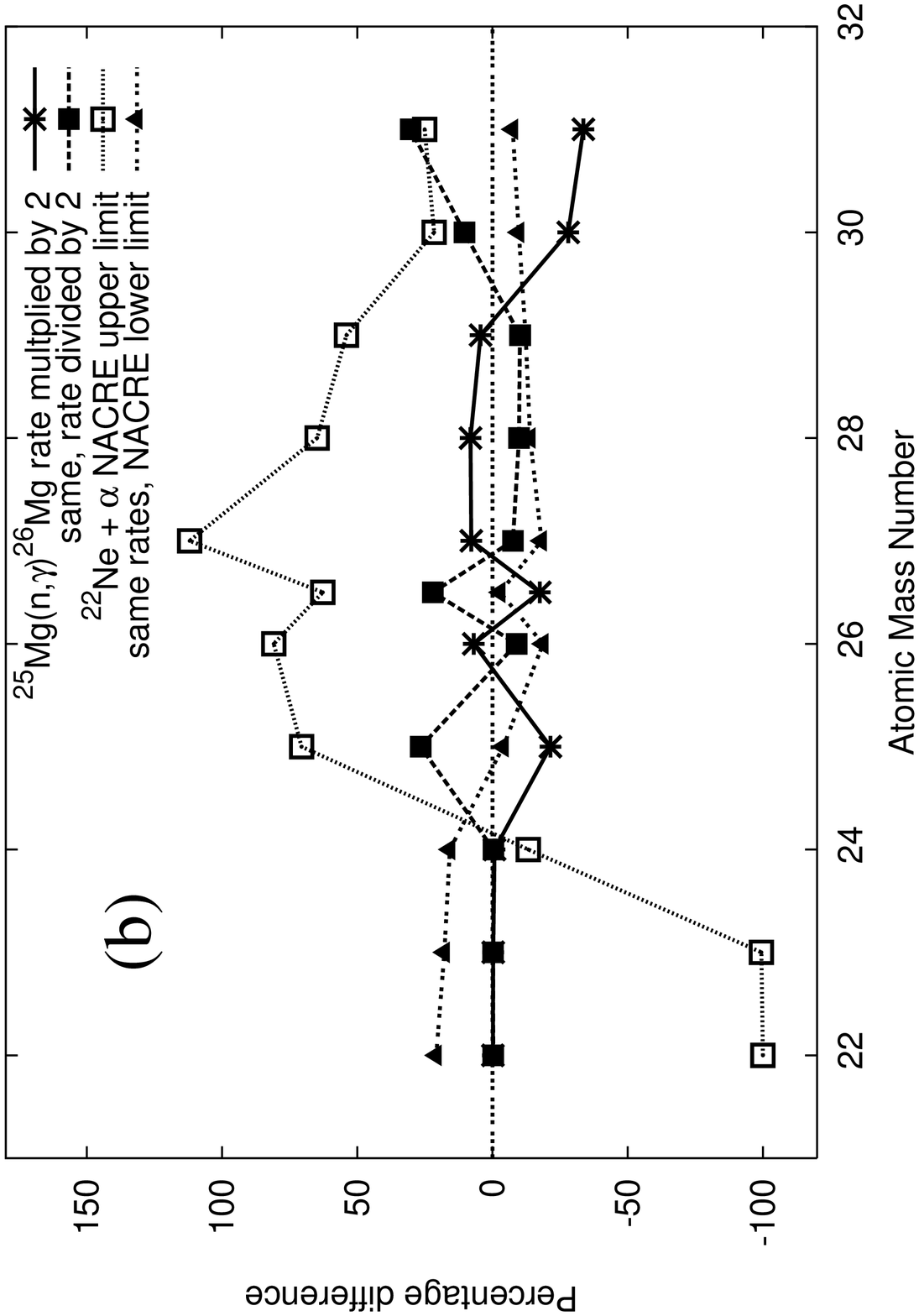}
\end{figure}

\clearpage

\begin{figure}
\plotone{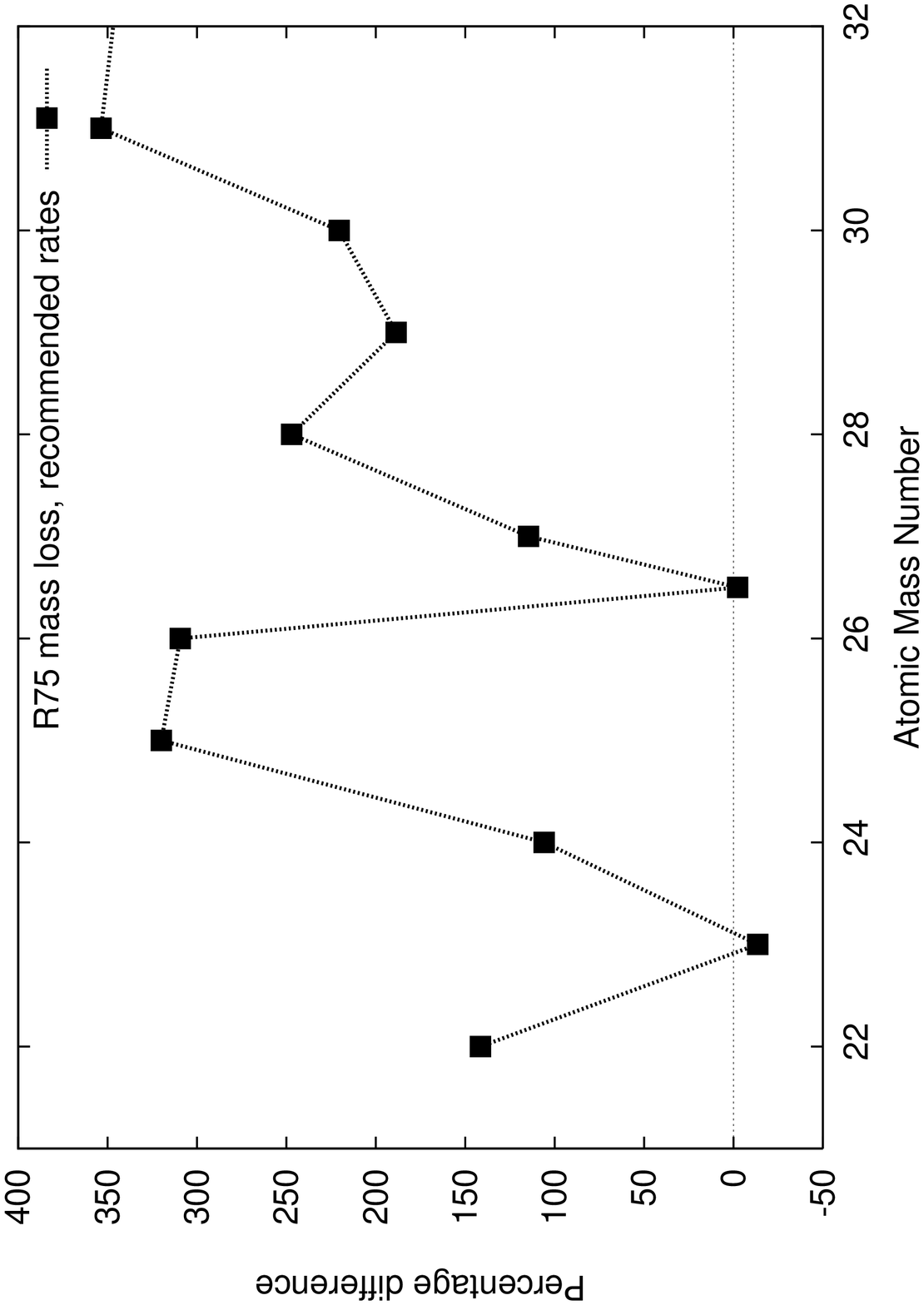} 
\end{figure}

\clearpage

\begin{figure}
\plotone{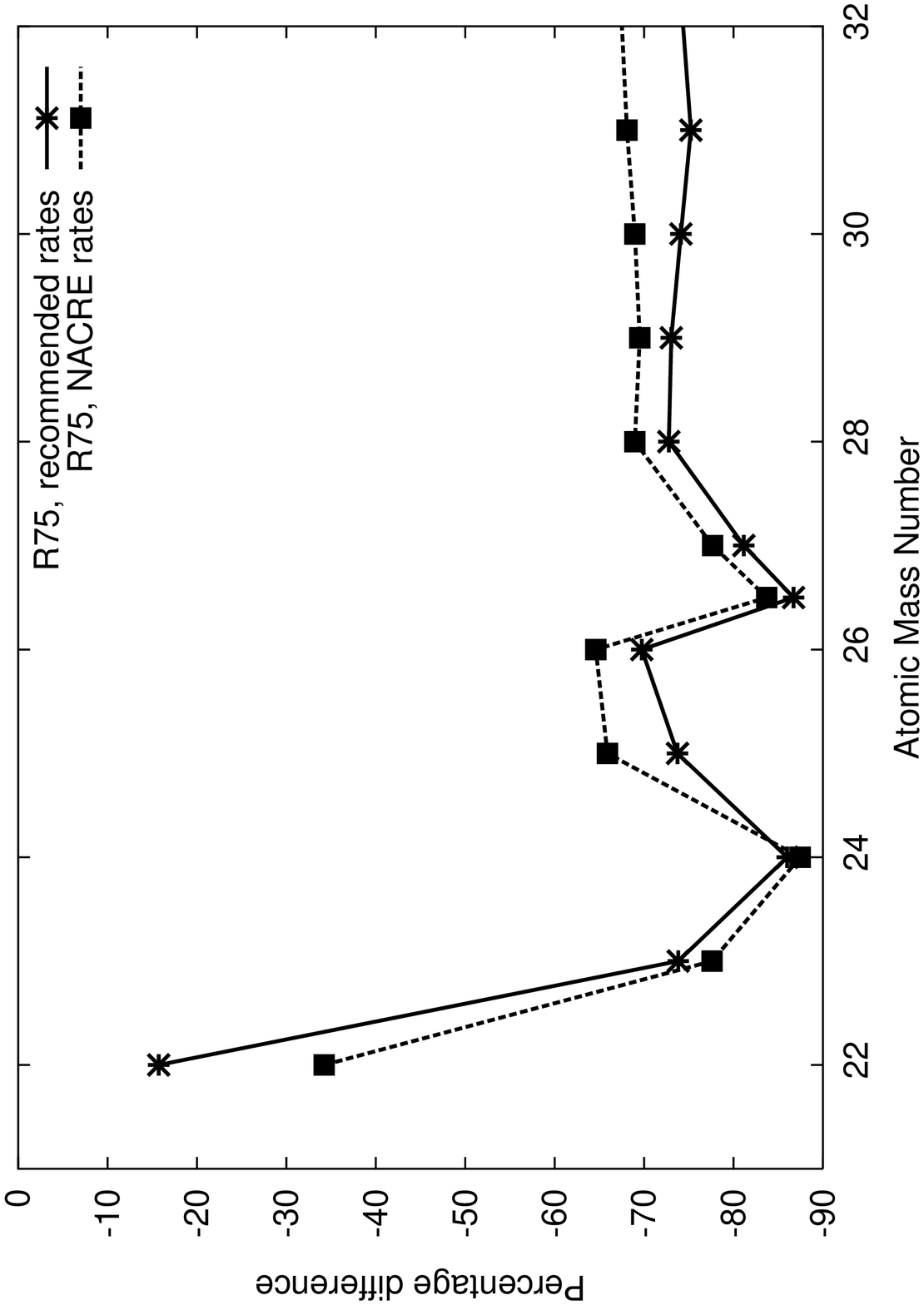} 
\end{figure}
\end{document}